\documentclass[superscriptaddress,showpacs,amssymb,10pt,reprint,aps,prd,longbibliography,nofootinbib,floatfix]{revtex4-2}

\usepackage{graphicx,epsfig,amssymb} 
\usepackage{amsmath,amsfonts, times}
\usepackage{bm} 

\usepackage[normalem]{ulem}

\usepackage{epstopdf}
\usepackage[linktocpage,colorlinks]{hyperref}
\usepackage[caption=false]{subfig}
\usepackage[usenames]{color}     
\usepackage{natbib}
\usepackage{soul}
\usepackage{subfig}
\usepackage[utf8x]{inputenc}

\definecolor{coolblack}{rgb}{0.0, 0.18, 0.39}
\definecolor{darkred}{rgb}{0.5,0,0}
\definecolor{darkgreen}{rgb}{0,0.5,0}
\definecolor{darkblue}{rgb}{0,0,0.5}
\definecolor{lapislazuli}{rgb}{0.15, 0.38, 0.61}
\definecolor{venetianred}{rgb}{0.78, 0.03, 0.08}
\definecolor{bleudefrance}{rgb}{0.19, 0.55, 0.91}
\definecolor{dogwoodrose}{rgb}{0.84, 0.09, 0.41}
\hypersetup{colorlinks=true, citecolor=darkgreen, linkcolor=darkblue, 
urlcolor = blue}

\def\be{\begin{equation}}
\def\ee{\end{equation}}

\newcommand\numberthis{\addtocounter{equation}{1}\tag{\theequation}}

\newcommand{\bea}{\begin{eqnarray}}
\newcommand{\eea}{\end{eqnarray}}
\newcommand{\ben}{\begin{enumerate}}
\newcommand{\een}{\end{enumerate}}
\newcommand{\bi}{\begin{itemize}}
\newcommand{\ei}{\end{itemize}}

\newcommand{\rt}{r_{\star}}

\def\ga{\mathrel{\raise.3ex\hbox{$>$\kern-.75em\lower1ex\hbox{$\sim$}}}}
\def\la{\mathrel{\raise.3ex\hbox{$<$\kern-.75em\lower1ex\hbox{$\sim$}}}}

\def\l{\left}
\def\r{\right}
\def\be{\begin{equation}}
\def\ee{\end{equation}}

\def\I_M{{I_{\scriptscriptstyle M\times M}}}

\def\be{\begin{equation}}
\def\ee{\end{equation}}
\def\bea{\begin{eqnarray}}
\def\eea{\end{eqnarray}}
\newcommand{\beq}{\begin{eqnarray}}
\newcommand{\eeq}{\end{eqnarray}}
\def\pa{\partial}

\begin{document}
\title{\large Scattering by deformed black holes}

\author{Renan B. Magalh\~aes}
\email{renan.magalhaes@icen.ufpa.br}
\affiliation{Programa de P\'os-Gradua\c{c}\~{a}o em F\'{\i}sica, Universidade 
	Federal do Par\'a, 66075-110, Bel\'em, Par\'a, Brazil.}

\author{Luiz C. S. Leite}
\email{luiz.leite@ifpa.edu.br}
\affiliation{Campus Altamira, Instituto Federal do Par\'a, 68377-630, Altamira, Par\'a, Brazil.}

\author{Lu\'is C. B. Crispino}
\email{crispino@ufpa.br}
\affiliation{Programa de P\'os-Gradua\c{c}\~{a}o em F\'{\i}sica, Universidade 
	Federal do Par\'a, 66075-110, Bel\'em, Par\'a, Brazil.}
\affiliation{Faculdade de F\'{\i}sica, Universidade 
Federal do Par\'a, 66075-110, Bel\'em, Par\'a, Brazil.}

\begin{abstract}
Obtaining black hole solutions in alternative theories of gravity can be a difficult task due to cumbersome field equations that arise in many of such theories. In order to study the strong field regime in a model-free approach, one can consider deformed black hole solutions with additional parameters beyond mass, charge and angular momentum. We investigate the scattering and absorption of a massless scalar field by non-Schwarzschild black holes, considering the Johannsen and Psaltis parametrization. In particular, we study the scattering process by the classical (null geodesics analysis), semiclassical (glory approximation) and partial waves approaches. We present the formalism needed to compute the scattering and absorption of massless scalar waves by Schwarzschild-like BHs with a set of deformation parameters in addition to their mass, and present a selection of our results. We compare our numerical results for the scattering cross section with the classical and semiclassical results, obtaining excellent agreement.  
\end{abstract}

\date{\today}

\maketitle

\section{Introduction}\label{sec:int}
Black holes (BHs) are among the most fascinating predictions of 20th century physics. They are defined by a one-way surface, called an event horizon, where not even light can return once it crosses this boundary. Within general relativity (GR), BH physics is rigorously constructed over the uniqueness and no-hair theorems~\cite{robinson2009,chrusciel2012stationary}, which establish that BHs are described by only three parameters, their mass, charge and angular momentum~\cite{chandrasekhar}. 

When considering modified gravity theories, we are also able to find BHs with nontrivial hairs~\cite{kanti1996,yunes2011} so that BHs can be seen as natural candidates to test corrections of GR~\cite{PMCC2011}. The recent experimental advances, which led to the observation of gravitational wave signals from binary BH systems~\cite{ligo} and to the first observation of a BH shadow~\cite{eht}, provide new channels to test gravity in the strong field regime, as well as to test the no-hair theorems~\cite{isi2019}.

An interesting approach to construct BH solutions regular on and outside the event horizon was proposed by Johannsen and Psaltis (JP)~\cite{JP2011}, which may be used in the analysis of the tests of gravity in the near BH horizon region, without considering a particular alternative theory of gravity. The JP approach is based on writing a BH line element containing deviations from a standard BH solution, which are called parametric deformations. These ``deformed'' BHs can, in principle, be associated to BH solutions in alternative theories of gravity by appropriate choices of the parameters, and when the deformation parameters vanish, the standard solutions are recovered.
 The most simple version of the JP BHs (JPBHs), with only one nonvanishing parameter, has been widely studied over the years, both in their rotating and nonrotating versions~\cite{B2011,KZ2013,P2013,Bambi2015,Pappas2019}, and many aspects of such spacetimes have been already investigated, like the possibility to support naked singularities~\cite{P2013}.
The JP approach has been followed by other parametrizations with different insights and motivations~\cite{CPR2014,LZ2014,KZ2016}.

The presence of matter and energy surrounding BHs is associated with interesting phenomena, which have been investigated both from the theoretical and from the experimental point of view. In this context, the study of the dynamics of fundamental fields in the vicinity of parameterized BHs naturally arises as a subject of interest. In particular, two processes, namely the absorption and scattering of fundamental fields by BHs, have been studied over the years considering different fields propagating in various BH spacetimes~\cite{fabbri,unruh,sanchez,BODC:2014,CDHO2014,CDHO:2015,LDC:2017,LDC:2017,LDC:2018,LBC:2019}. More recently, the absorption of massless scalar waves by parameterized BHs has been investigated considering two distinct parametrized static solutions~\cite{MLC:2020PLB,MLC:2020EPJC}. 

We investigate the scattering of massless scalar waves by a static JPBH, by considering the classical (null geodesics analysis), the semiclassical (glory approximation) and the partial waves approaches, in order to understand how the deformation parameter influences the scattering process, and we compute the differential scattering cross section via these three approaches. We obtain the fundamental equations to analyze the scattering problem for a static JPBH with an arbitrary number of parameters, and show our results for the simplest case (with a single nonvanishing parameter). We compare the three approaches, finding, e. g., an excellent agreement between the semiclassical and the partial waves description for large values of the scattering angle. 

The remaining of this paper is organized as follows: In Sec.~\ref{sec:non-schw}, we briefly present and discuss some properties of the static JPBH. In Sec.~\ref{sec:class}, we consider the trajectories of massless particles in the vicinity of static JPBHs, obtaining the main equations used to describe null geodesics scattered by these objects. In Sec.~\ref{sec:partialwave}, we analyze the massless scalar field dynamics in the JPBH spacetime, presenting the partial waves approach and giving the expressions of the absorption and scattering cross sections. Additionally, we present the glory approximation, valid for high frequencies and large scattering angles. 
A selection of our numerical results is presented in Sec.~\ref{sec:results}. We conclude with our final remarks in Sec.~\ref{sec:remarks}. Throughout the paper, we use natural units $G = c = \hbar = 1$ and signature $(+ − −\, −)$.

\section{Non-Schwarzschild spacetime}\label{sec:non-schw}
We consider nonrotating BHs with parametric deformations, whose line element is given by~\cite{JP2011} 
\be
\label{eq:non_schw_ln}
ds^2=h(r)\left(f(r)dt^2-\frac{1}{f(r)}dr^2\right)-r^2\left(d\theta^2+\sin^2\theta d\varphi^2\right),
\ee
where $f(r)=1-2M/r$ and $h(r)$ is chosen to be
\begin{equation}
\label{eq:def_h}
h(r)\equiv 1+\sum_{n=0}^{\infty}\epsilon_n\left(\dfrac{M}{r}\right)^n.
\end{equation} 
When all of the so-called deformation parameters $\epsilon_n$ vanish, we obtain Schwarzschild's spacetime.
The line element~\eqref{eq:non_schw_ln} describes a static BH with an arbitrary number of parameters; hence, it may violate the no-hair theorems.
 
Requiring Eq.~\eqref{eq:non_schw_ln} to be asymptotically compatible with the Schwarzschild line element, $\epsilon_0$ and $\epsilon_1$ must vanish. Moreover, for the line element~\eqref{eq:non_schw_ln} to agree with experiments related to the Einstein's equivalence principle~\cite{boggs}, the parameter $\epsilon_2$ must be constrained by $\vert\epsilon_2\vert\leq 4.6\times 10^{-4}$~\cite{JP2011}, and henceforth, we assume $\epsilon_2=0$.

The location of the event horizon is $r_{h} = 2M$, independently of the parameters present in the JPBH, coinciding with the horizon location in the standard Schwarzschild solution.
This implies that the area of a JPBH horizon is independent of the deformation parameters, being equal to the area of a Schwarzschild BH with the same mass parameter $M$.


By considering only the first non-null term $\epsilon_3$ in the sum present in Eq.~\eqref{eq:def_h}, the line element~\eqref{eq:non_schw_ln} reduces to 
\begin{align*}
\label{eq:def_h_3}
ds^2=&\left[1+\epsilon_3\left(\dfrac{M}{r}\right)^3\right]\left(f(r)dt^2-\frac{1}{f(r)}dr^2\right)\\&-r^2\left(d\theta^2+\sin^2\theta d\varphi^2\right). \numberthis
\end{align*}

It is important to point out that for positive values of the parameter $\epsilon_3$, the line element~\eqref{eq:def_h_3} has only one singularity, located at $r=0$, while for negative values of $\epsilon_3$, the line element has an additional singularity at $r=|\epsilon_3|^{1/3}M$~\cite{KZ2013,MLC:2020EPJC}. If the parameter $\epsilon_3$ is sufficiently negative, namely $\epsilon_3\leq-8$, the singularity lies outside ($\epsilon_3<-8$) or on ($\epsilon_3=-8$) the event horizon. Along this paper, we consider only BH solutions regular on and outside the event horizon, and henceforth, we assume $\epsilon_3>-8$.

\section{Classical approach: absorption and scattering of lightlike trajectories}
\label{sec:class}
In this section, we obtain the main equations used to analyze the lightlike trajectories in the static parameterized spacetimes described by Eq.~\eqref{eq:non_schw_ln}. 
%
In such spacetimes, the geodesics satisfy
\begin{equation}
\label{eq:sph_geodesics}
\dot{s}^2=h(r)f(r)\dot{t}^2-\dfrac{h(r)}{f(r)}\dot{r}^2-r^2\left(\dot{\theta}^2+\sin^2\theta \dot{\varphi}^2\right)=k,
\end{equation}
where the overdot represents the derivative with respect to an affine parameter. The kinetic term $\dot{s}^2\equiv g_{\mu\nu}\dot{x}^{\mu}\dot{x}^{\nu}$ is normalized to 1 for massive particles and has zero norm for massless particles, i.e., $k=1$ for timelike trajectories and $k=0$ for lightlike (null-like) trajectories. Following from the stationarity and spherical symmetry of the spacetime, we have two conserved quantities along the geodesics, namely 
\begin{align}
\label{eq:E} \dot{t}\,h(r)f(r) &= E,\\
\label{eq:L} \dot{\varphi}\,r^2\sin^2\theta &= L,
\end{align}
where $L$ is related to the angular momentum of the particle and $E$ to its energy. Due to the spherical symmetry,
one can restrict the geodesic analysis to the equatorial plane ($\theta=\pi/2$) without loss of generality~\cite{chandrasekhar}.

Inserting Eqs.~\eqref{eq:E} and~\eqref{eq:L} into Eq.~\eqref{eq:sph_geodesics}, one finds, for massless particles~($k=0$),
\begin{equation}
\label{eq:orbit_equation}
\dfrac{h(r)f(r)}{r^2}\left[\dfrac{h(r)}{r^2f(r)}\left(\dfrac{dr}{d\varphi}\right)^2+1\right]-\dfrac{1}{b^2}=0,
\end{equation}
where $b\equiv L/E$ is the impact parameter. After differentiating Eq.~\eqref{eq:orbit_equation} with respect to $\varphi$, we obtain
\begin{align*}
2r[h(r)]^2\dfrac{d^2r}{d\varphi^2}+2h(r)\left(\dfrac{dr}{d\varphi}\right)^2\left[rh'(r)-2h(r)\right]\\+r^2\left\{r[h(r)f(r)]'-2[h(r)f(r)]\right\}=0, \numberthis \label{eq:eq2r}
\end{align*}
where the prime denotes a differentiation with respect to $r$. Equation~\eqref{eq:eq2r} describes the lightlike trajectories in the spacetimes with line element~\eqref{eq:non_schw_ln}. By solving Eq.~\eqref{eq:eq2r} using appropriate boundary conditions, we fully determine the null-like geodesics. Moreover, with Eq.~\eqref{eq:eq2r}, we can find a light ring at $r=r_c$, defined by $\ddot{r}=0$ and $\dot{r}=0$, or equivalently $d^2r/d\varphi^2=0$ and $dr/d\varphi=0$. The light ring location $r_c$ can be determined by solving
\begin{equation}
\label{eq:light_sphere}
r[h(r)f(r)]'-2[h(r)f(r)]=0.
\end{equation} 
Substituting $r_c$ into Eq.~\eqref{eq:orbit_equation} we find the critical impact parameter $b_c$, such that the light rays (null geodesics) incoming from infinity with $b>b_c$ are scattered by the BH, whereas null geodesics with impact parameter $b<b_c$ are absorbed by the BH. Light rays with impact parameter $b=b_c$ stay in an unstable orbit at $r=r_c$. Geodesics with impact parameter sufficiently close to the critical one ($b\approx b_c$) may undergo several turns around the BH before being scattered ($b	\gtrsim b_c$) or absorbed ($b\lesssim b_c$).

\begin{figure}[h!]%
	\includegraphics[width=\columnwidth]{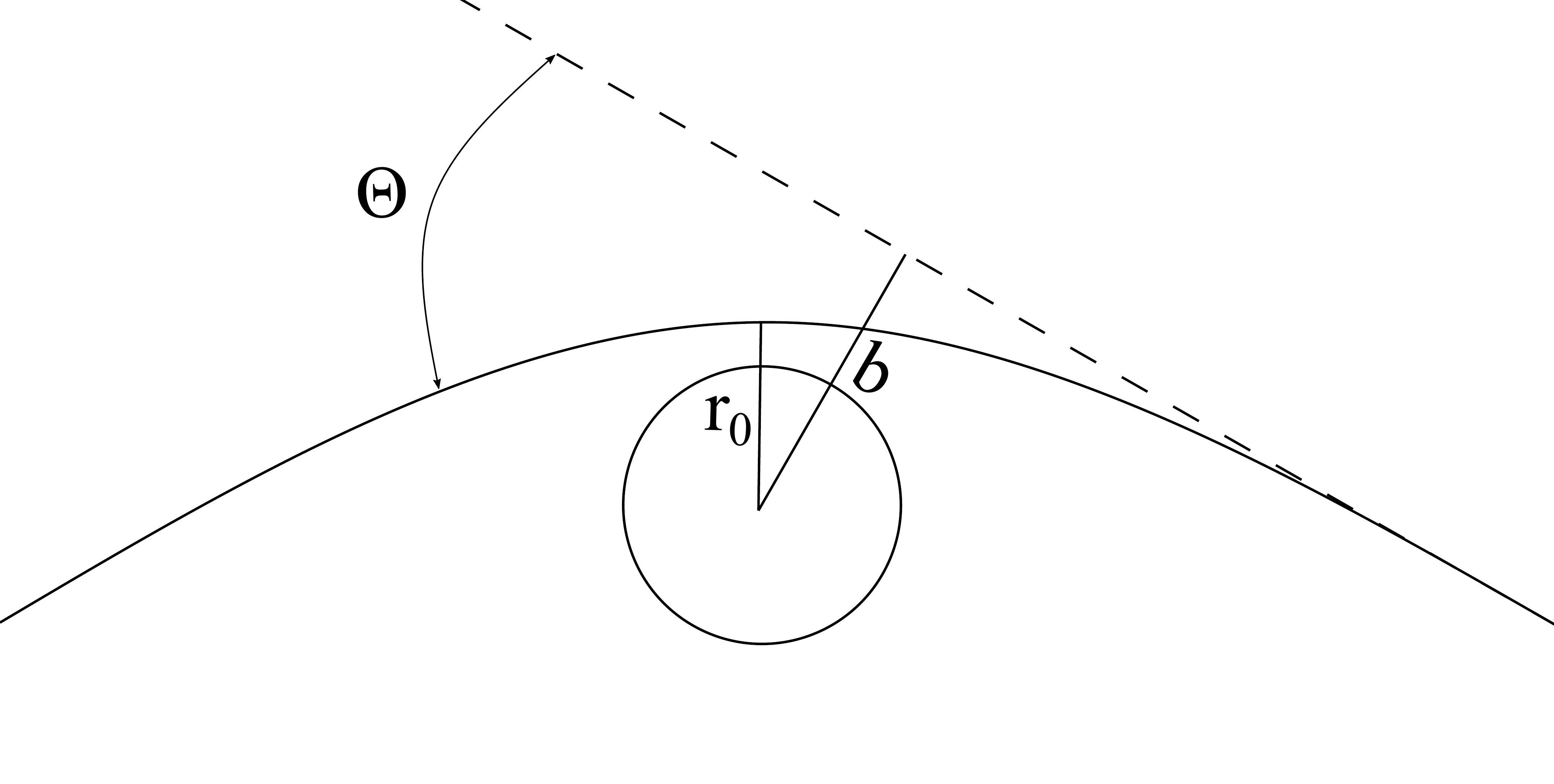}
	\caption{Schematic representation of a deflected trajectory. $\Theta$ is the total deflection angle, $b$ is the impact parameter and $r_0$ is the  minimum radial distance.}%
	\label{fig:light_trajectory}%
\end{figure}

In Fig.~\ref{fig:light_trajectory}, we present a deflected trajectory due the presence of a compact object. Far away from the compact object, the trajectory is a straight line, since we are assuming an asymptotically flat spacetime. The particle comes from infinity with an impact parameter $b>b_c$; close to the object, its path is bent, reaching a minimum radial distance, $r_0$, from the object, and then it is scattered back to infinity with a total deflection angle $\Theta$.

By integrating Eq.~\eqref{eq:orbit_equation}, one finds for massless particles~\cite{Weinberg},
\begin{equation}
\label{eq:phi_r}
\varphi(r)\, -\,\varphi_{\infty} = \int_{r}^{\infty} d\bar{r}\dfrac{r_0}{\bar{r}}\dfrac{h(\bar{r})}{\sqrt{\bar{r}^2 h(r_0)f(r_0)-r_0^2h(\bar{r})f(\bar{r})}},
\end{equation}
where $\varphi_{\infty}$ is the incidence direction of the light ray, and we have used
\begin{equation}
\label{eq:b_r0}
b^2 = \dfrac{r^2_0}{h(r_0)f(r_0)},
\end{equation}
since $(dr/d\varphi)=0$ at $r_0$. The total variation in $\varphi$ is simply $2|\varphi(r_0)-\varphi_{\infty}|$, and
\begin{equation}
\label{eq:deflection_angle}
\Theta(b)\equiv 2\left|\int_{r_{0}}^{\infty}dr\dfrac{b}{r}\dfrac{h(r)}{\sqrt{r^2-b^2h(r)f(r)}}\right|-\pi
\end{equation}
is the total deflection angle.

By considering incident null geodesics with impact parameter $b$, one can derive the classical scattering cross section, given by the following infinite sum~\cite{collins}:
\begin{equation}
\label{eq:class_scatt_cross}
\dfrac{d\sigma_{sc}}{d\Omega} = \dfrac{1}{\sin\theta}\sum_{n} b(\theta)\left|\dfrac{db}{d\theta}\right|,
\end{equation}
where $\theta=|\Theta-2n\pi|$ is the scattering angle, and $n=0,1,2,\dots$ tells us how many loops the massless particle undergoes around the compact scattering center before being scattered. 
The relation between the impact parameter $b$ and the scattering angle $\theta$ can be obtained by inverting Eq.~\eqref{eq:deflection_angle}. 

Additionally, in BH scenarios, one can define the classical absorption cross section (also called geometrical absorption cross section) by the area of a disk with radius $b_c$, namely
\begin{equation}
\label{eq:class_abs_cross}
\sigma_{abs}\equiv \pi b_{c}^2.
\end{equation}

\section{Partial waves approach: absorption and scattering of lightlike trajectories}\label{sec:partialwave}
\subsection{Massless scalar linear perturbations} \label{sec:scalarfield}
In order to investigate the scattering of massless scalar waves by static parameterized BH solutions, we consider a massless scalar field governed by the (massless) Klein-Gordon equation,
\be
(-g)^{-1/2}\pa_\mu(g^{\mu\nu}\sqrt{-g}\pa_\nu\Psi)=0,\label{eq:kge}
\ee
with $\pa_\mu\equiv\pa/\pa x^\mu$, $g^{\mu\nu}$ being the contravariant components of the metric and $g$ being the metric determinant, which, from Eq.~\eqref{eq:non_schw_ln}, reads
\begin{equation}
\label{eq:det_g}
g=-\left[r^2h(r)\sin\theta\right]^2.
\end{equation}

The massless scalar field~$\Psi$ can be conveniently decomposed as follows:
\be
\Psi(x^\mu)=\frac{\psi_{\omega l}(r)}{r}Y_{lm}(\theta,\,\varphi)e^{-i\omega t},
\label{eq:decom}
\ee
where $Y_{lm}$ are the spherical harmonics, which satisfy the eigenvalue equation 
\begin{equation}
\label{eq:Ylm}
\dfrac{1}{\sin\theta}\dfrac{\partial}{\partial\theta}\left(\sin\theta\dfrac{\partial Y_{lm}}{\partial\theta}\right)+\dfrac{1}{\sin^2\theta}\dfrac{\partial^2Y_{lm}}{\partial\varphi^2}=-l(l+1)Y_{lm}.
\end{equation}
By using the decomposition 
~\eqref{eq:decom}, 
we end up with
the following ordinary differential equation for the radial function $\psi_{\omega l}$:
\be
f(r)\frac{d}{dr}\l[f(r)\frac{d\psi_{\omega l}}{dr}\r]+\l[\omega^2-V_l(r)\r]\psi_{\omega l}=0,\label{eq:radialeq}
\ee
where $V_l$ is the effective potential, given by
\be
V_l(r)\equiv \frac{f(r)}{r}\frac{df}{dr}+f(r)h(r)\frac{l(l+1)}{r^2}.\label{eq:effective_potential} 
\ee
We notice that all dependence on the additional parameters $\epsilon_n$ is contained  in the term in the effective potential which also depends on the angular momentum index $l$. This implies that, for the lower angular mode $l=0$, the radial solution $\psi_{\omega 0}$ is the same for any JPBH and coincides with the one of the Schwarzschild BH case.

By introducing the tortoise coordinate $\rt$, implicitly defined by
\begin{equation}
\label{eq:tortoise}
\dfrac{d\rt}{dr} \equiv \dfrac{1}{f(r)},
\end{equation}
we can rewrite Eq.~\eqref{eq:radialeq} as a Schrödinger-like equation, being free of first order derivatives, namely
\be
\frac{d^2\psi_{\omega l}}{d\rt^2}+\l[\omega^2-V_l(\rt)\r]\psi_{\omega l}(\rt)=0.\label{eq:radialeqtor}
\ee
From Eq.~\eqref{eq:effective_potential}, we notice that the effective potential vanishes at the event horizon, $r=r_h$ $(\rt\rightarrow-\infty)$, and at the spatial infinity, $r\rightarrow\infty$ $(\rt\rightarrow\infty)$. 
Considering this asymptotic behavior of the effective potential and assuming incoming waves from infinity, the corresponding solution of Eq.~\eqref{eq:radialeqtor} satisfies the following boundary conditions:
\be
\psi_{\omega l}(\rt)\sim\l\{
\begin{array}{ll}
	e^{-i \omega \rt}+{\cal R}_{\omega l }e^{i \omega \rt},&\rt\to+\infty, \\
	{\cal T}_{\omega l }e^{-i\omega \rt},& \rt\to-\infty,
\end{array}\r.\label{eq:inmodes}
\ee
where the coefficients ${\cal R}_{\omega l }$ and ${\cal T}_{\omega l }$ can be related to the reflection and transmission coefficients, respectively. Moreover, they satisfy the flux conservation relation,
\be
|{\cal R}_{\omega l }|^2+|{\cal T}_{\omega l }|^2=1.
\ee

\subsection{Scattering and absorption cross sections} \label{sec:cross_sections}
In this section, we present the main steps to investigate  the total absorption cross section and the differential scattering cross section of a massless scalar field impinging on a spherically symmetric BH.

\subsubsection{Absorption cross section}
The total absorption cross section, $\sigma_{\text{abs}}$, can be defined as the ratio between the flux of $\Psi$ through the BH horizon  and the current of the incident wave. 
Using the decomposition of the scalar field into partial waves, $\sigma_{\text{abs}}$ can be written as a sum of the partial absorption cross sections, $\sigma^{(l)}_{\text{abs}}$, namely~\cite{MC:2014, BC:2016}
%
\begin{equation}
\sigma_{\text{abs}} = \sum_{l=0}^{\infty}\sigma^{(l)}_{\text{abs}},
\end{equation}
and the $\sigma^{(l)}_{\text{abs}}$ are given in terms of the coefficient ${\cal T}_{\omega l }$ as
\begin{equation}
\sigma^{(l)}_{\text{abs}} = \dfrac{\pi}{\omega^2}(2l+1)|{\cal T}_{\omega l }|^2.
\end{equation}
For static spacetimes, in the low- and high-frequency regimes, the behavior of the total absorption cross section is well described by some analytical results. In the low-frequency regime, the total absorption cross section is equal to the area of the event horizon~\cite{das, Higuchi:2001}, while in the high-frequency regime, the total absorption cross section oscillates around the geometrical absorption cross section~\eqref{eq:class_abs_cross}~\cite{folacci}.

\subsubsection{Scattering cross section}
The scattering amplitude $\hat{g}(\theta)$ is given by~\cite{CDO,MOC:2015}
\begin{equation}
\label{eq:partial_wave_series}
\hat{g}(\theta) = \dfrac{1}{2i\omega}\sum_{l=0}^{\infty}(2l+1)\left[(-1)^{l+1}{\cal R}_{\omega l }-1\right]P_l(\cos\theta),
\end{equation}
such that the differential scattering cross section is
\begin{equation}
\dfrac{d\sigma_{sc}}{d\Omega} \equiv |\hat{g}(\theta)|^2.
\end{equation}
The scattering amplitude $\hat{g}(\theta)$ has, in general, a poor convergence. Actually it is required an infinite number of partial waves to describe the divergence of $\hat{g}(\theta)$ at $\theta=0$. In order to handle with it, one can consider the series reduction method, first introduced to the electron scattering problem~\cite{YRW:1954} and generalized to BH scenarios (cf., e.g., Ref.~\cite{SLDC:2020}), or equivalently, by considering the Complex Angular Momentum approach, which is based on a sum over the Regge poles of the $S$ matrix~\cite{FH:2019}. To obtain the results shown in Sec. \ref{sec:results}, we used the series reduction procedure discussed in Ref.~\cite{SLDC:2020}. 

At large scattering angles, just as in optics, there is a glory effect associated to the massless scalar field scattering by a BH. The glory is a bright spot or halo in the backward direction, whose magnitude size and brightness may be obtained by the semiclassical approximation introduced in Ref.~\cite{matzner}, namely
\begin{equation}
\label{eq:glory}
\dfrac{d\sigma_{sc}}{d\Omega}\Big\vert_{\theta\approx\pi} \approx 2\pi \omega b^2_{g}\left\vert\dfrac{db}{d\theta}\right\vert_{\theta\approx\pi} [J_{0}(\omega b_g \sin\theta)]^2,
\end{equation}
where $b(\pi)\equiv b_g$ is the glory impact parameter, and $J_{0}(x)$ is the Bessel function of the first kind of order $0$. The approximation~\eqref{eq:glory} is valid in the high-frequency regime ($\omega M\gg 1$). However, as we shall see in Sec.~\ref{sec:results}, even for $\omega M \sim 1$, this approximation presents good agreement with the numerical results at $\theta\sim\pi$.

The existence of the photon ring, which enables light to be deflected through arbitrarily large angles, allows different values of $b_g$ (each of them implying in a different total deflection angle $\Theta$) to be related to the same scattering angle. For instance, the multiple values $\Theta=\pi+2n\pi$ (with $n\in\mathbb{N}$ being the number of revolutions around the BH) result in backscattered rays, contributing to the glory effect. However, the most significant contribution comes from the $n=0$ case.

\section{Results}\label{sec:results}
In this section, we present our numerical results associated to the JPBHs and compare them with the Schwarzschild case. We consider JPBHs with only one deformation parameter different from zero, namely $\epsilon_3$. To simplify the notation, we adopt the redefinition $\epsilon_3\equiv\epsilon$. 

\subsection{Classical limit}\label{sec:classical_limit}
\begin{figure}[!h]
	\includegraphics[width=\columnwidth]{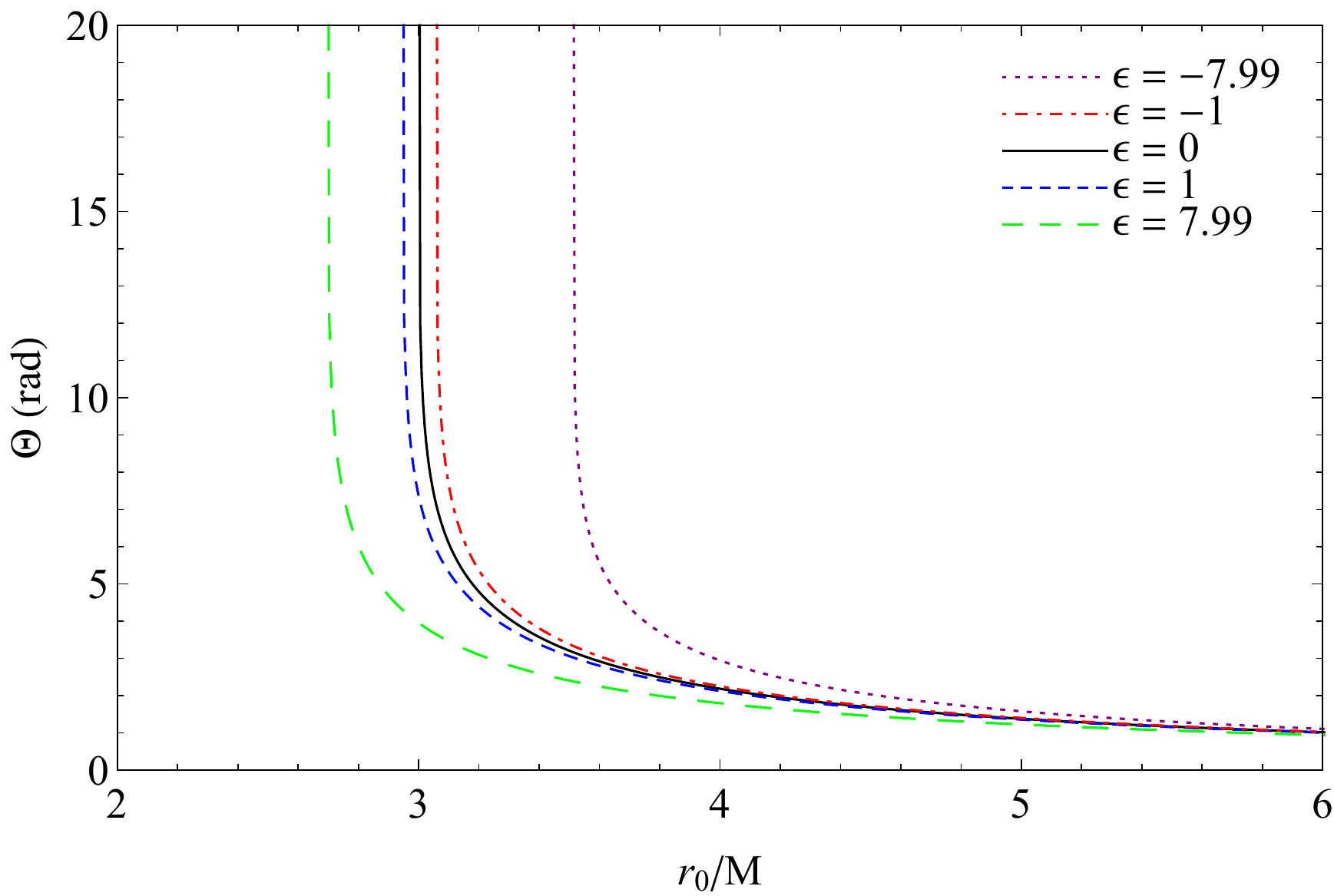}
	\caption{Total deflection angle (in radians) as a function of $r_0$ for some choices of the JPBH deformation parameter. }
	\label{fig:deflection angle}%
\end{figure}

\begin{figure}[!h]%
	\includegraphics[width=\columnwidth]{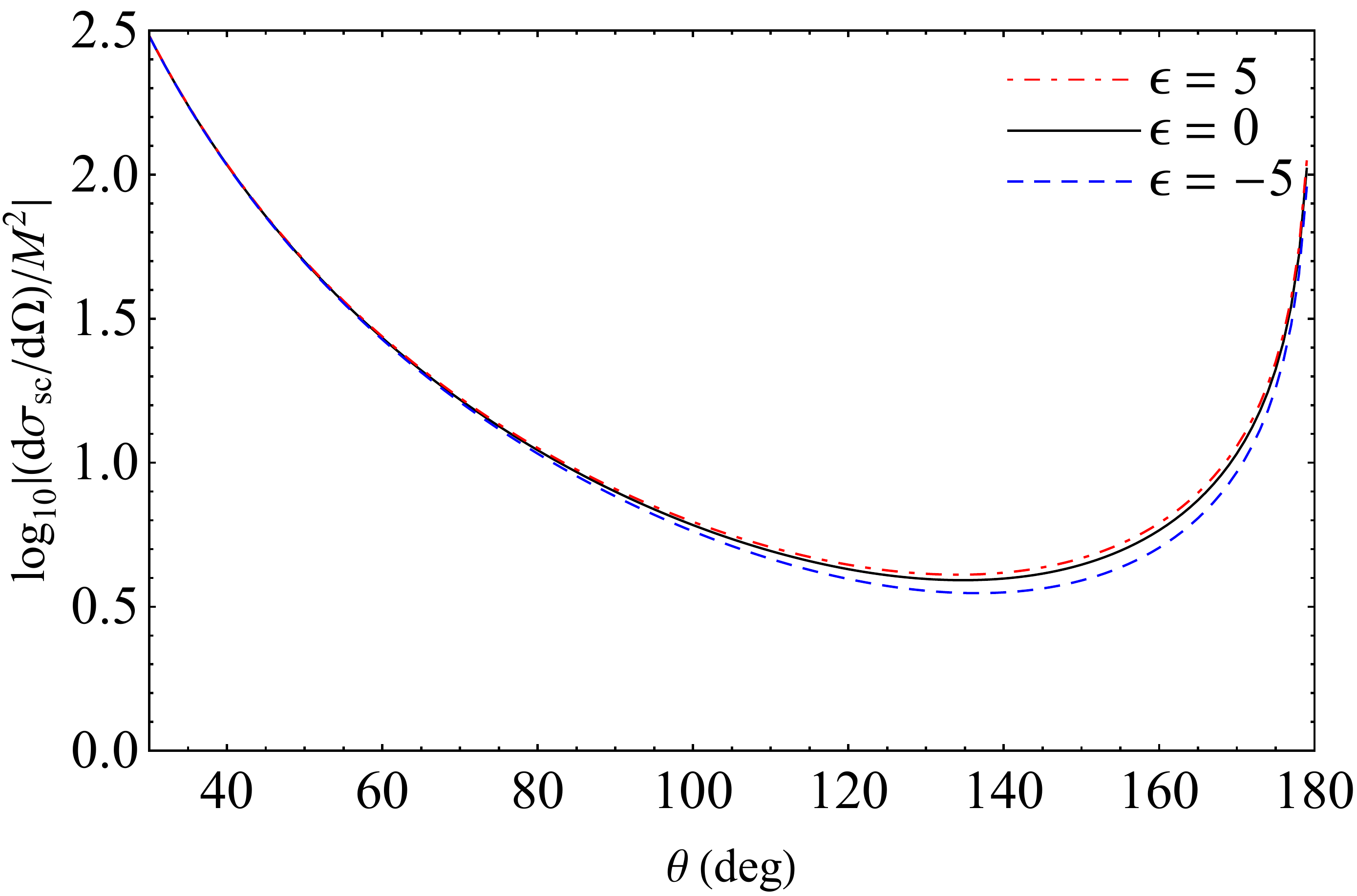}
	\caption{Comparison between the classical differential scattering cross section of JPBHs with the Schwarzschild one.}%
	\label{fig:class_scatt}%
\end{figure}

\begin{figure}[!h]%
	\includegraphics[width=\columnwidth]{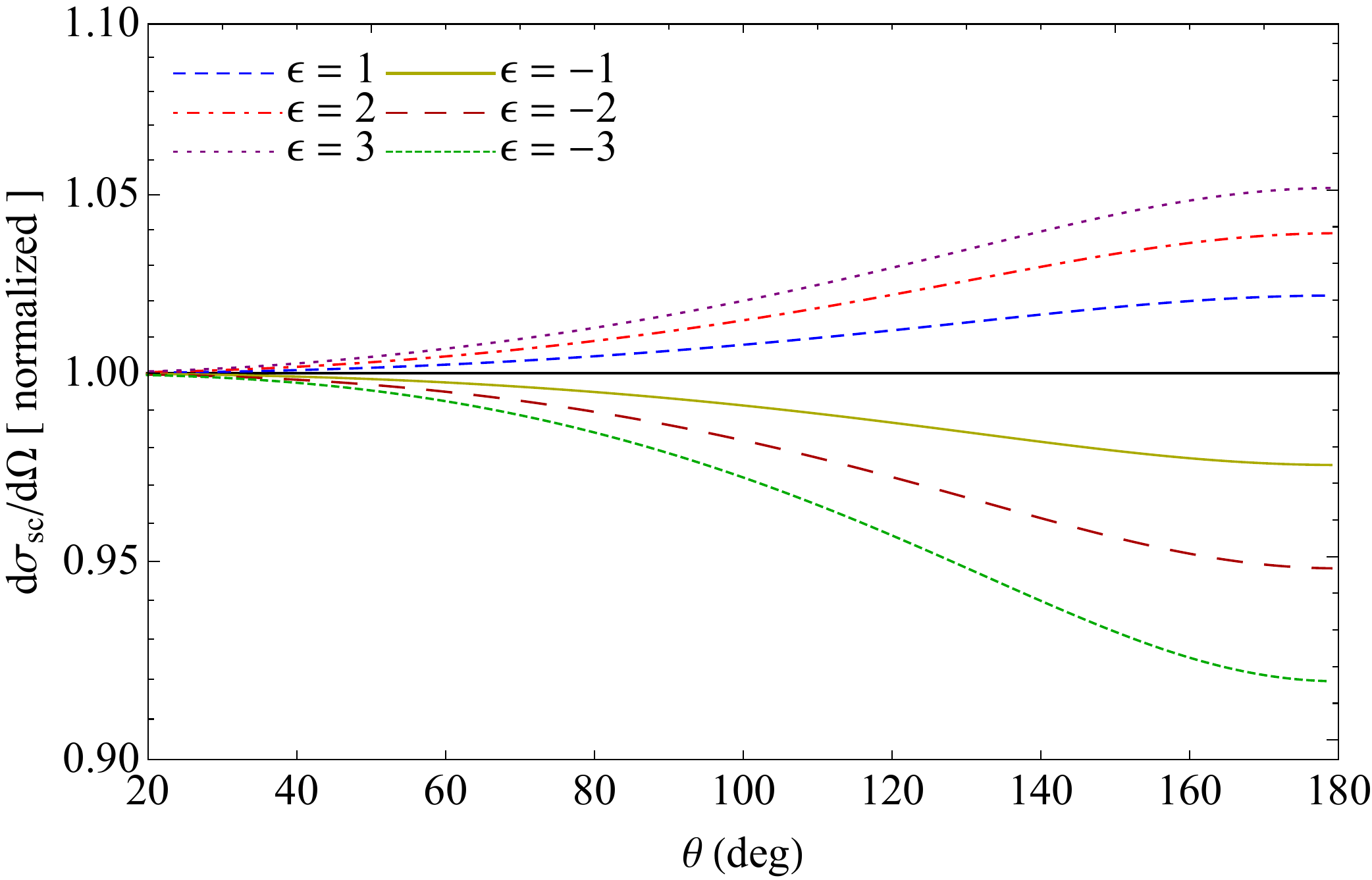}
	\caption{Classical scattering cross section of JPBHs, normalized (at each scattering angle) by the corresponding Schwarzschild value.}%
	\label{fig:class_scatt_normalized}%
\end{figure}

In Fig.~\ref{fig:deflection angle}, we plot the total deflection angle as a function of the closest radius for some choices of the deformation parameter of JPBHs. The vertical asymptotes of each curve are the unstable photon orbit radii $r_c$, which become smaller as we increase the value of the parameter $\epsilon$.
We notice that for near-critical orbits ($r_0\approx r_c$), the geodesics can be deflected by angles arbitrarily greater than $2\pi$, meaning that the geodesics may undergo many loops around the BH before being scattered.
In Fig.~\ref{fig:class_scatt}, we plot the classical scattering cross section obtained by Eq.~\eqref{eq:class_scatt_cross}, comparing two JPBHs, one with a positive value of $\epsilon$ ($\epsilon=5$) and another one with a negative value of $\epsilon$ ($\epsilon=-5$), with the Schwarzschild BH case ($\epsilon=0$). We notice that the scattering cross section, in the mid-to-large range of the scattering angle, is bigger (smaller) for positive (negative) values of the deformation parameter $\epsilon$, when compared to the Schwarzschild case ($\epsilon=0$). This behavior can be better observed in Fig.~\ref{fig:class_scatt_normalized}, where we plot the differential scattering cross section normalized by the corresponding Schwarzschild value, for some choices of the parameter $\epsilon$.

\subsection{Semiclassical limit}
\begin{figure}[!h]
\includegraphics[width=\columnwidth]{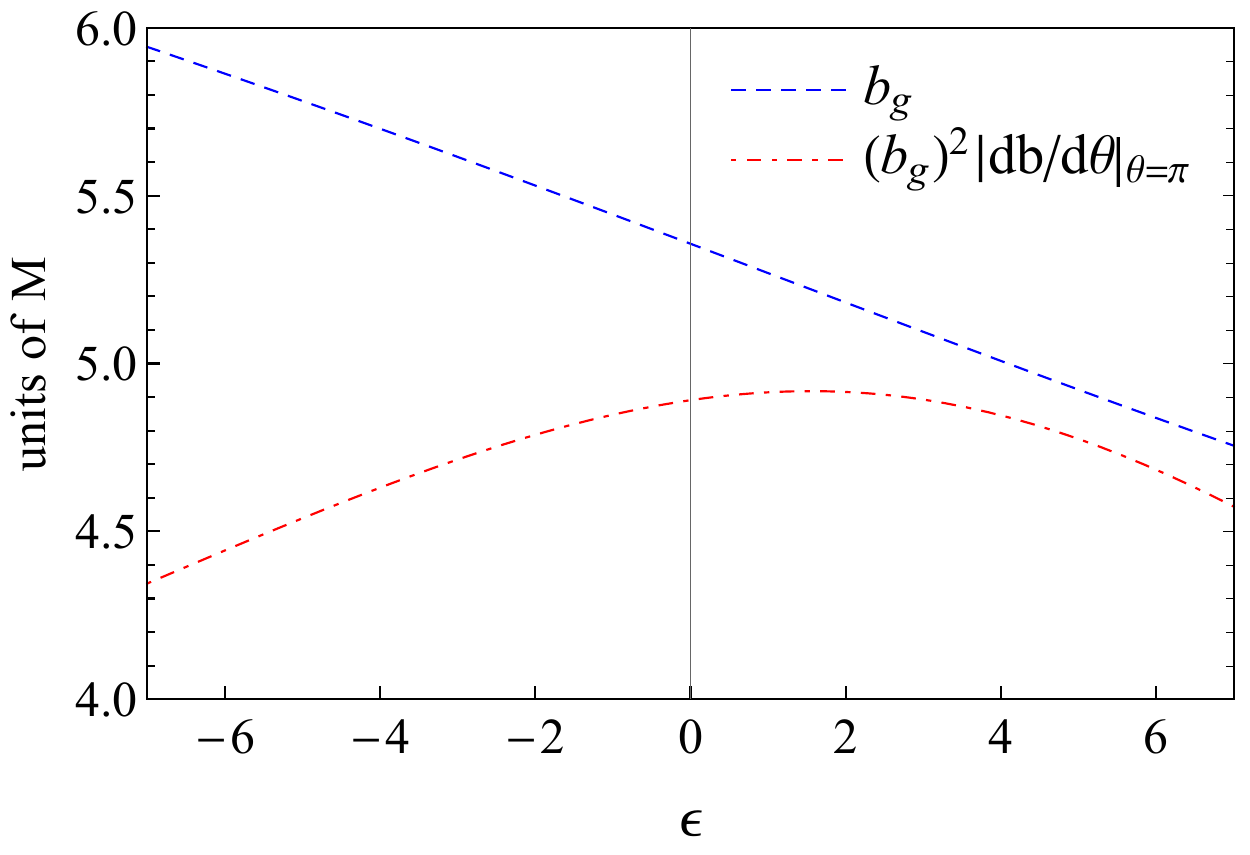}
\caption{Glory scattering parameters for JPBHs as a function of $\epsilon$, considering only the contribution from $n=0$. We notice that $b_g$ decreases monotonically with $\epsilon$, while $b_g^2|db/d\theta|_{\theta=\pi}$ has a local maximum at $\epsilon\approx 1.55$.}%
\label{fig:gloryparameters}%
\end{figure}

In Fig.~\ref{fig:gloryparameters}, we plot the relevant parameters for the glory scattering, namely $b_g$ and $b^2_g|db/d\theta|_{\theta = \pi}$ [see Eq.~\eqref{eq:glory}], as functions of the deformation parameter $\epsilon$. We notice from Fig.~\ref{fig:gloryparameters} that the glory impact parameter $b_g$ decreases monotonically as the deformation parameter $\epsilon$ increases.\footnote{A similar behavior occurs for charged BHs, such as the Reissner-Nordstr\"om BHs~\cite{CDO} and the Bardeen BHs~\cite{MOC:2015}, by increasing the BH charge.} Through the analysis of the argument of the Bessel function in Eq.~\eqref{eq:glory}, we notice that the interference fringe widths of the scattering cross section are proportional to $1/b_g$. Since $b_g$ decreases as $\epsilon$ increases, the fringes get narrower for smaller values of $\epsilon$. We also see from Fig.~\ref{fig:gloryparameters} that $b^2_g|db/d\theta|_{\theta = \pi}$ presents a local maximum at $\epsilon\approx 1.55$.
\subsection{Partial waves}
\subsubsection{Absorption}\label{sec:absorption_results}
\begin{figure}[!h]%
	\includegraphics[width=\columnwidth]{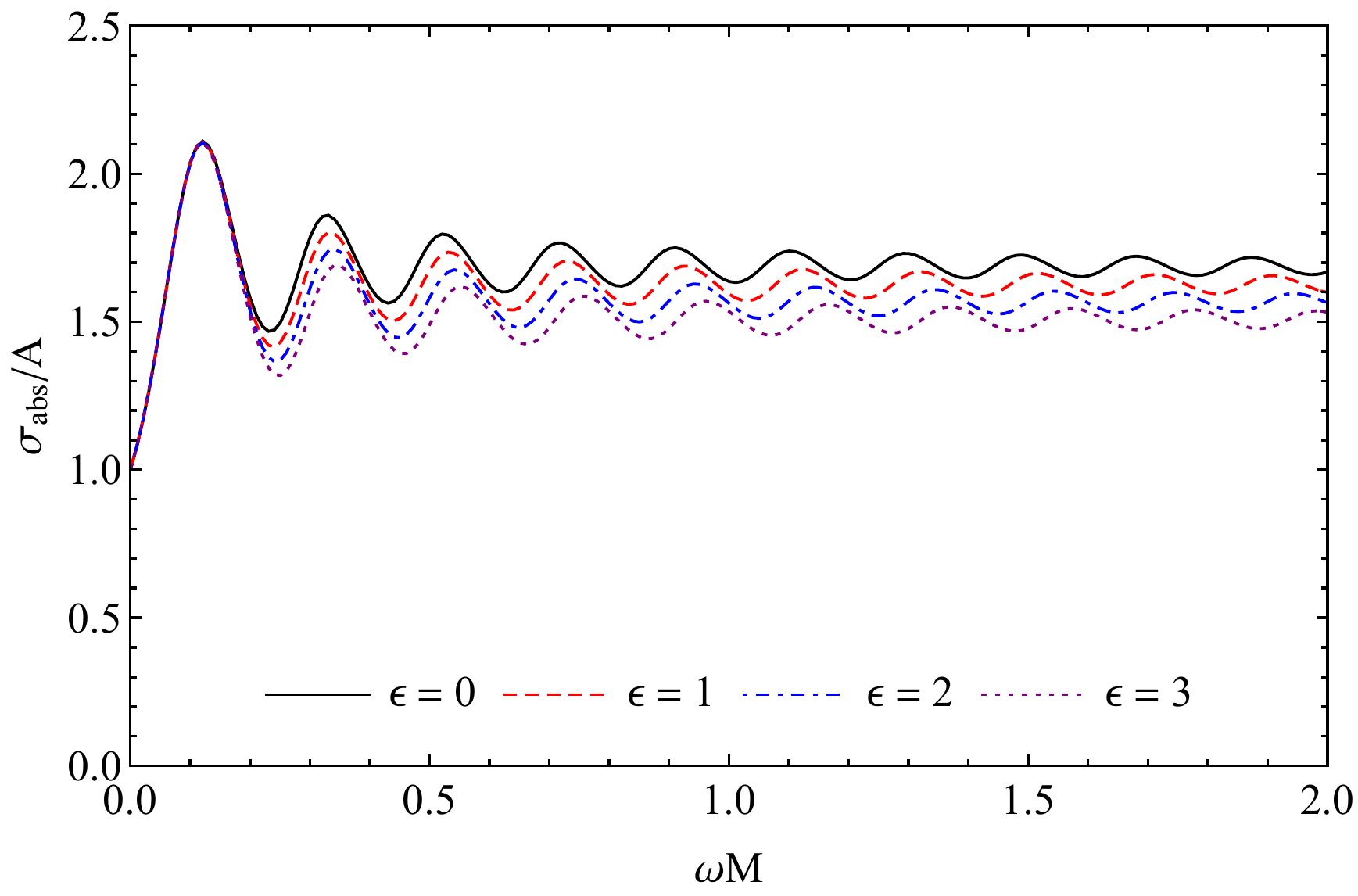}
	\includegraphics[width=\columnwidth]{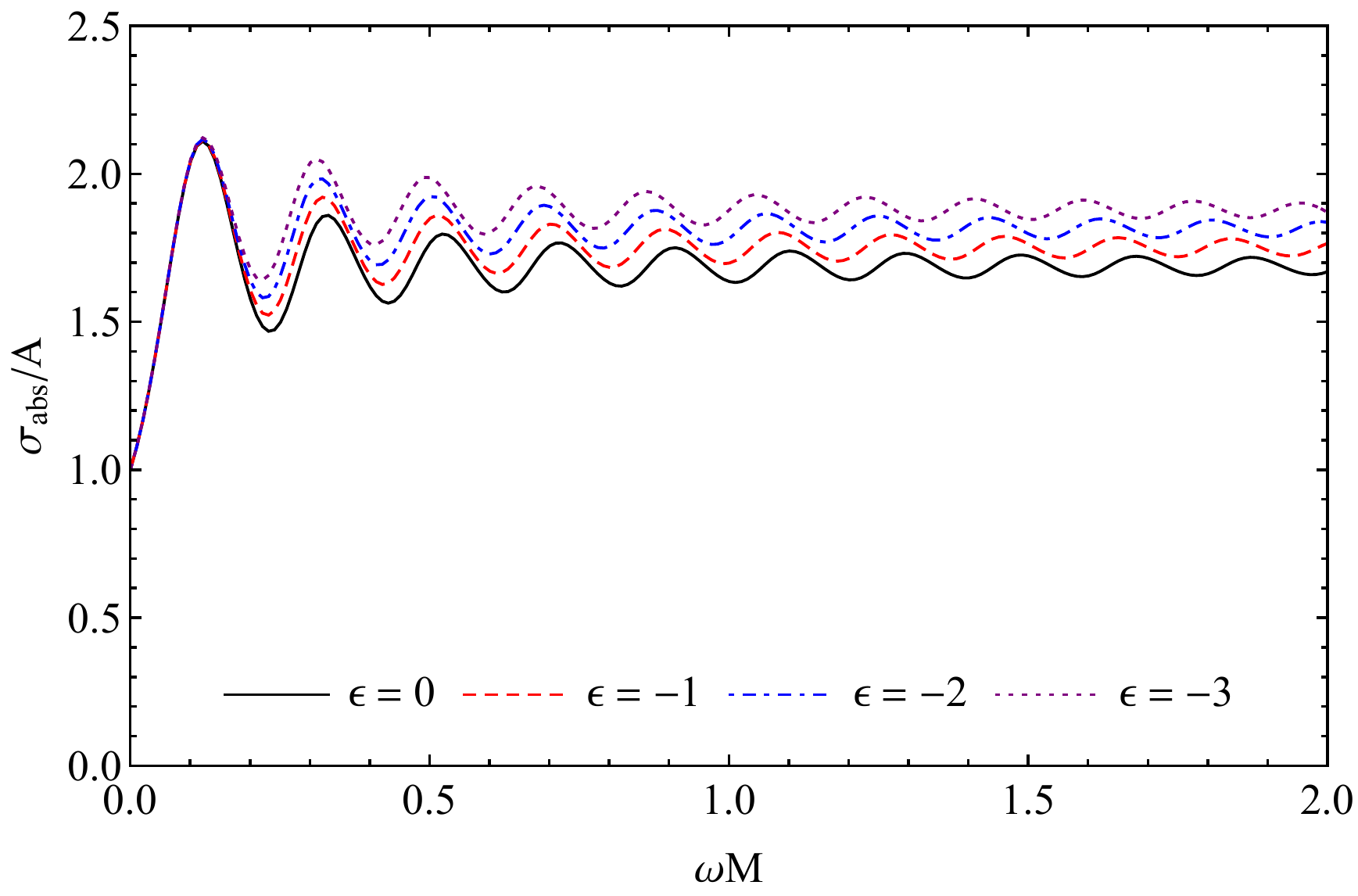}
		\caption{Total absorption cross section for JPBHs with non-negative (top panel) and nonpositive (bottom panel) values of $\epsilon$. We point out that the absorption diminishes as we increase the value of the deformation parameter.}%
\label{fig:abs_JP}%
\end{figure}

In Fig.~\ref{fig:abs_JP}, we plot the total absorption cross section for JPBHs divided by the horizon area, for non-negative (top panel) and nonpositive (bottom panel) values of $\epsilon$. In the low-frequency regime, the dominant contribution comes from the $l=0$ mode~\cite{MLC:2020EPJC}, and the total absorption cross section goes to the area of the JPBH, as expected by the general result of Refs.~\cite{das, Higuchi:2001}. In the mid-to-high-frequency regime, the total absorption cross section oscillates around the geometrical absorption cross section~\eqref{eq:class_abs_cross}. This behavior can also be obtained through the so-called sinc approximation~\cite{folacci}. We also notice from Fig.~\ref{fig:abs_JP} that the absorption is smaller for larger values of the deformation parameter. 
\subsubsection{Scattering}\label{sec:scattering_results}
\begin{figure}%
	\includegraphics[width=\columnwidth]{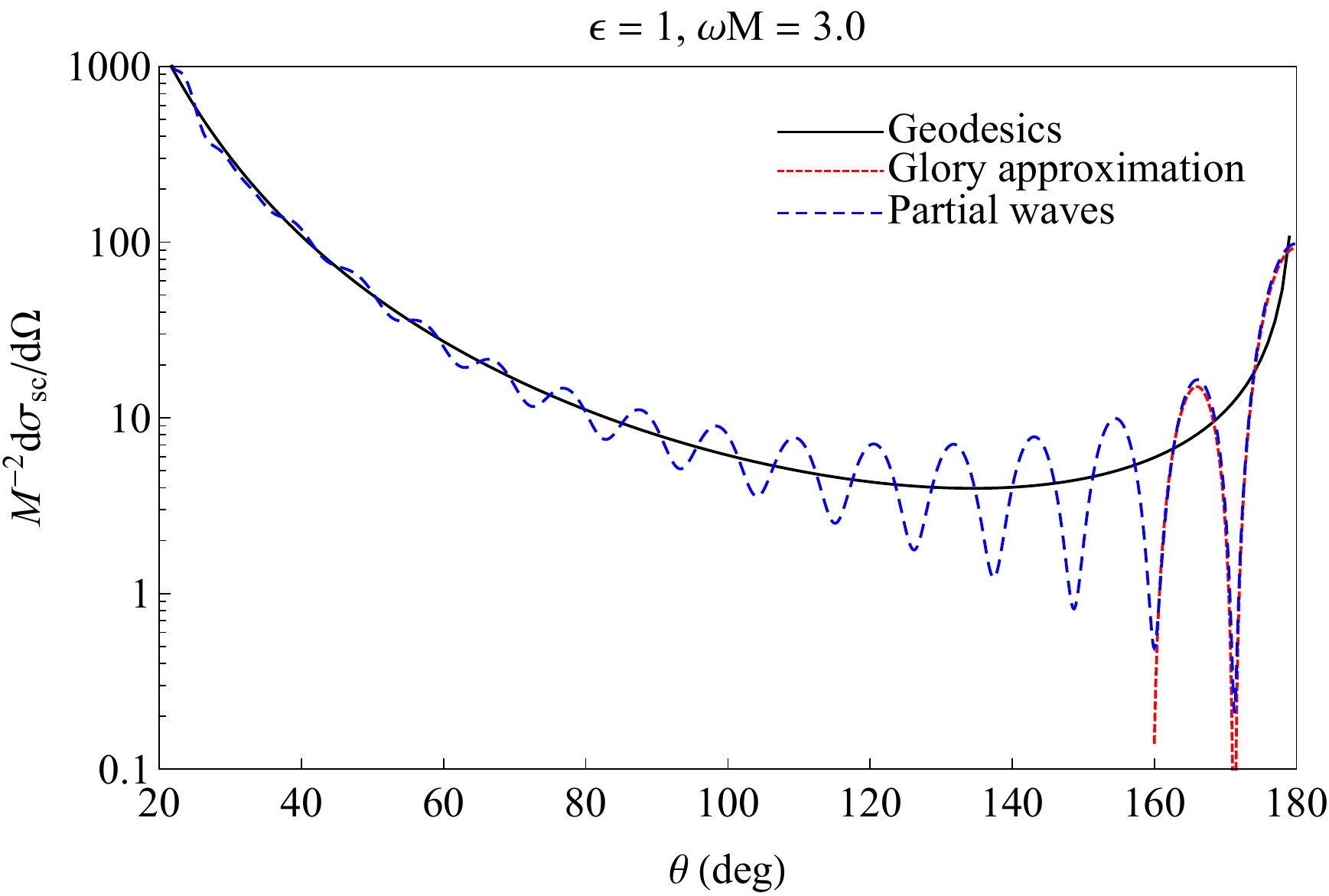}\\
	\includegraphics[width=\columnwidth]{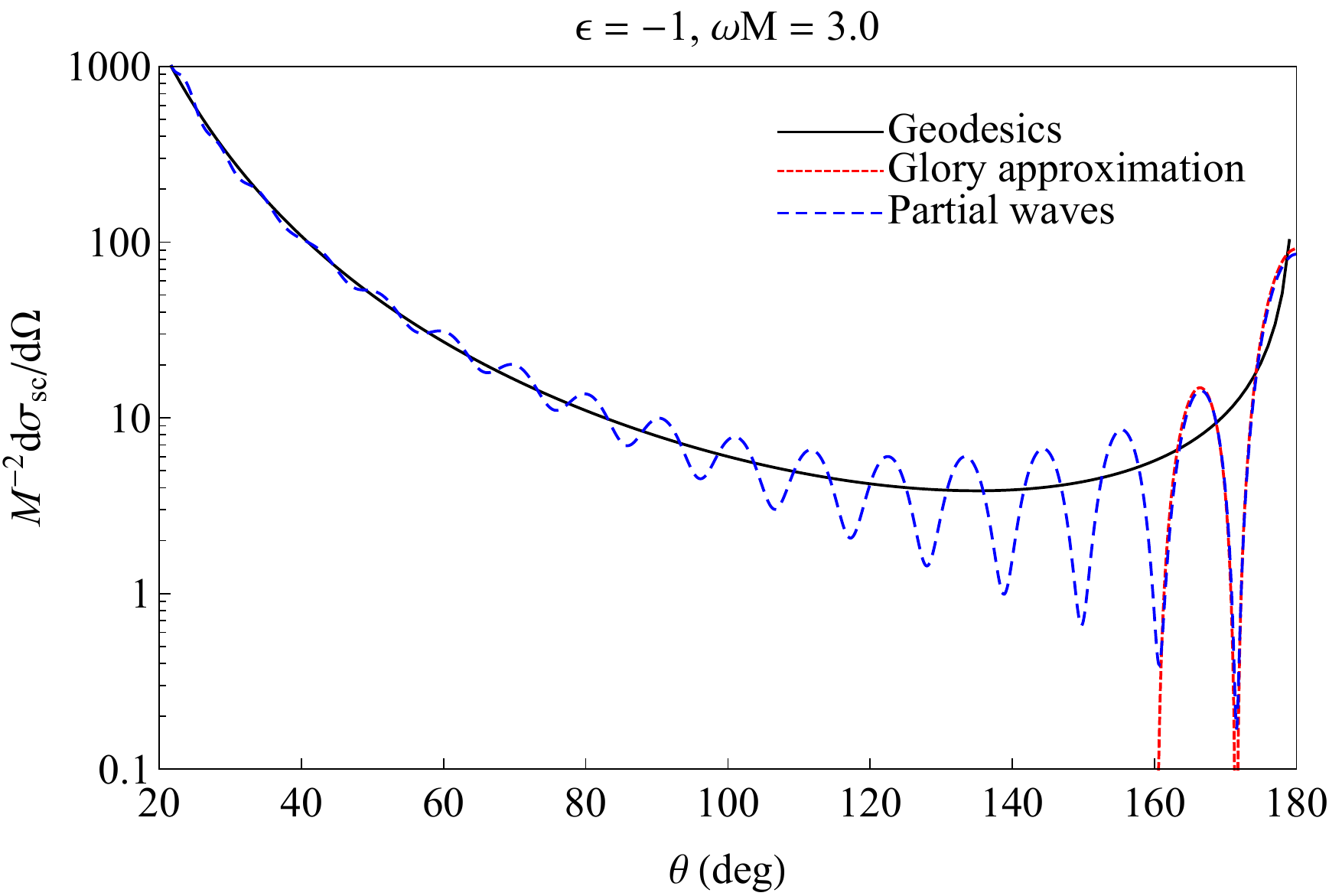}
	\caption{Comparison between the differential scattering cross section of JPBHs obtained by geodesics analysis, glory approximation and partial waves approach. We consider $\omega M = 3.0$ and $\epsilon=1$ (top panel) and $\epsilon=-1$ (bottom panel).}%
	\label{fig:class_glory_partial}%
\end{figure}

In Fig.~\ref{fig:class_glory_partial}, we plot the differential scattering cross section of two JPBHs, namely for $\epsilon=1$ (top panel) and $\epsilon=-1$ (bottom panel) values of the deformation parameter. In both panels, we assume $\omega M=3.0$. For comparison, we also plot the classical and glory (differential) scattering cross sections. Due to the long-range character of the gravitational interaction, the differential scattering cross section diverges in the forward direction~($\theta= 0$). From Fig.~\ref{fig:class_glory_partial}, we notice that the differential scattering cross section oscillates around the classical result. We also notice that the semiclassical glory approximation for $\theta\approx\pi$ shows excellent agreement with the numerical results. Moreover, since we are considering a scalar field, the scattered flux presents a maximum in the backward direction~($\theta=\pi$). 

\begin{figure*}
\centering
\subfloat{\label{sfig:a}\includegraphics[width=\columnwidth]{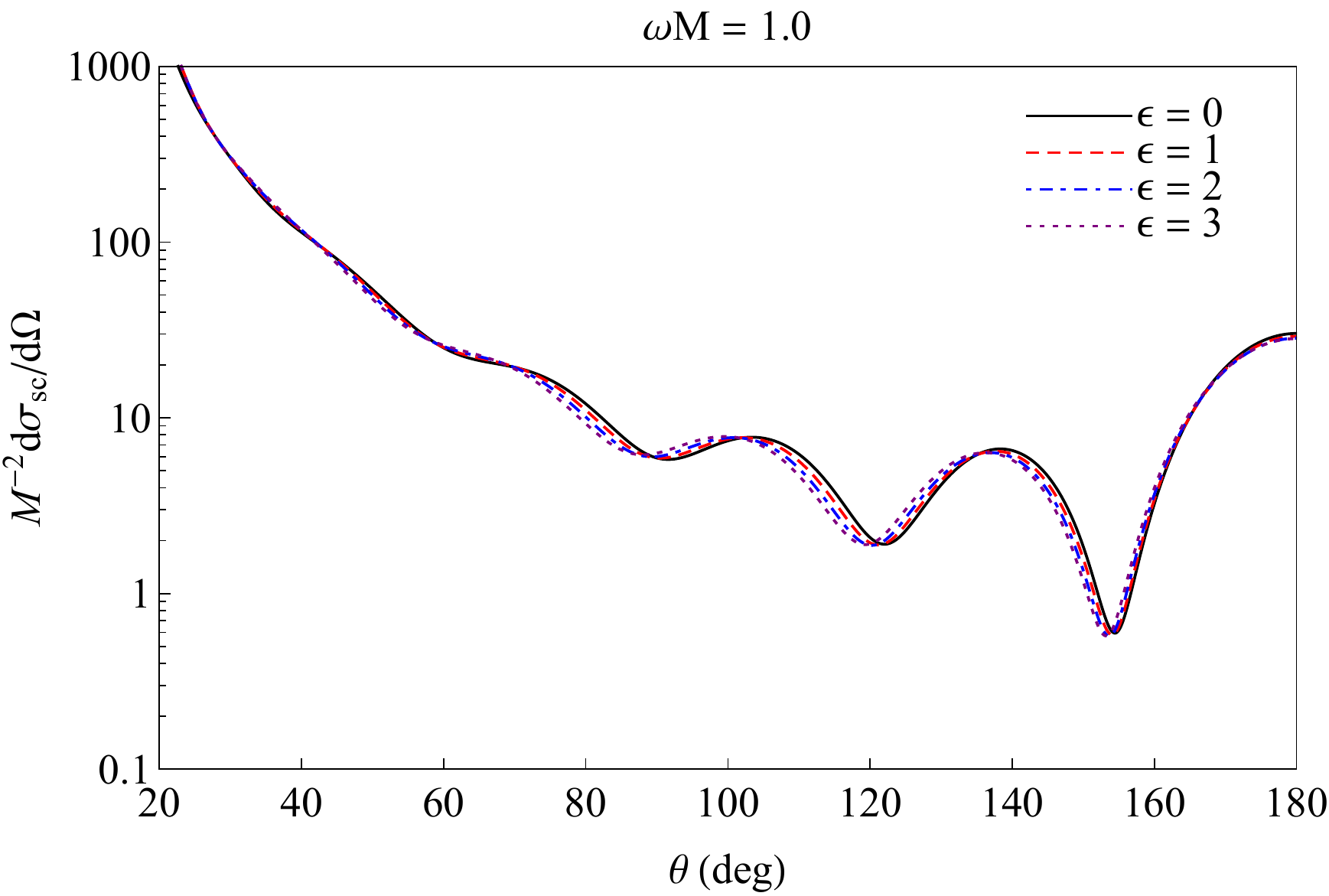}}
\subfloat{\label{sfig:b}\includegraphics[width=\columnwidth]{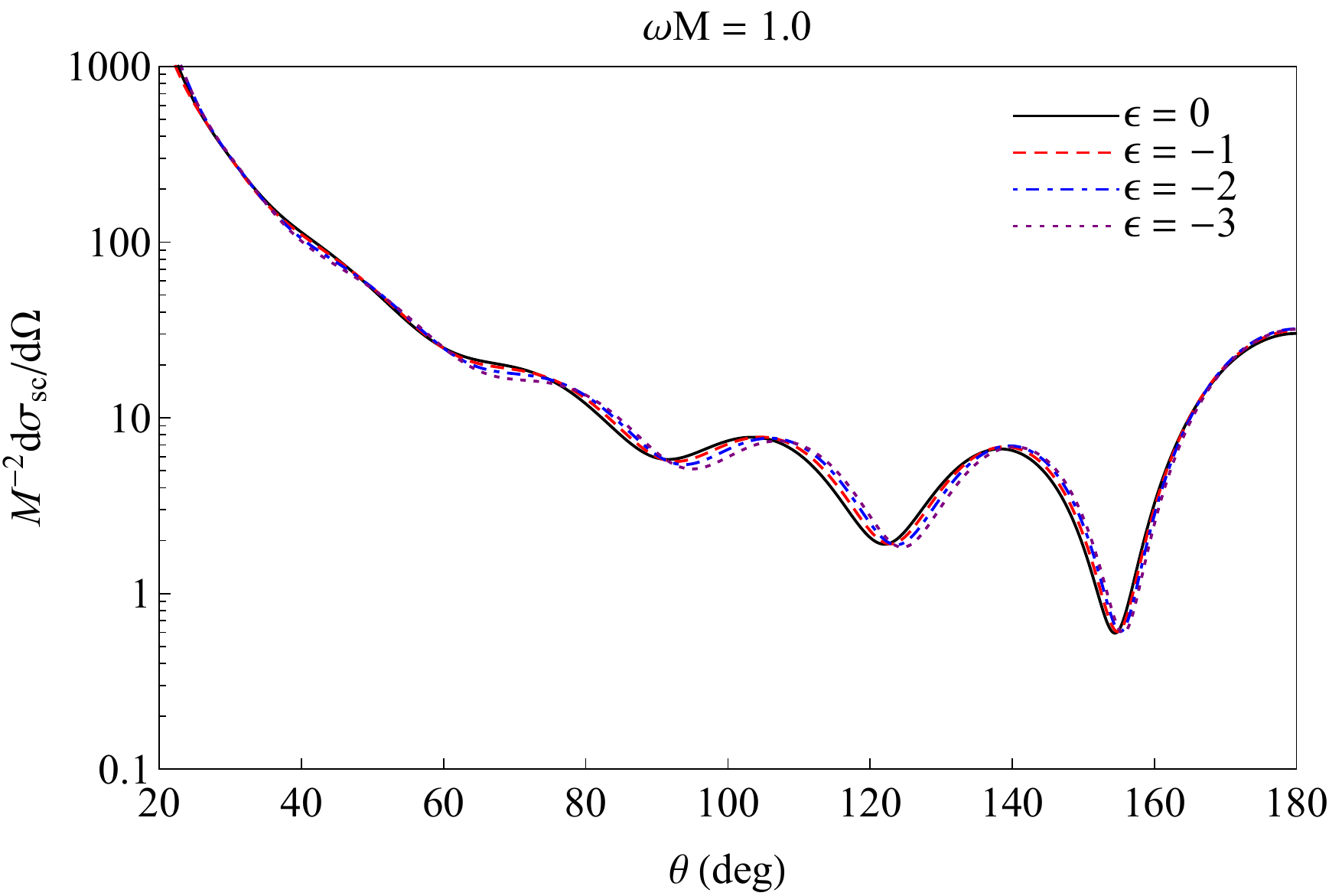}}\\
\subfloat{\label{sfig:a1}\includegraphics[width=\columnwidth]{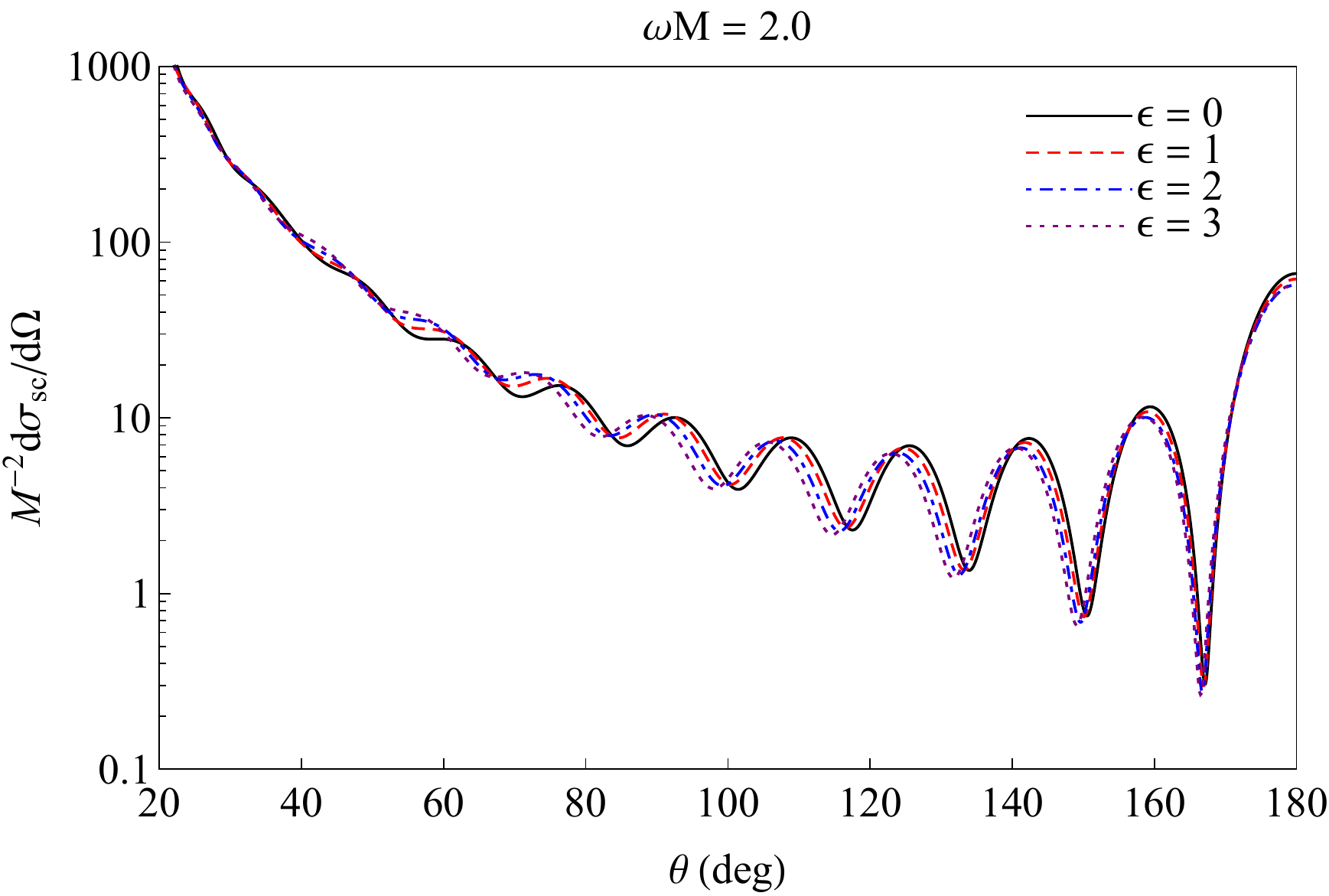}}
\subfloat{\label{sfig:b1}\includegraphics[width=\columnwidth]{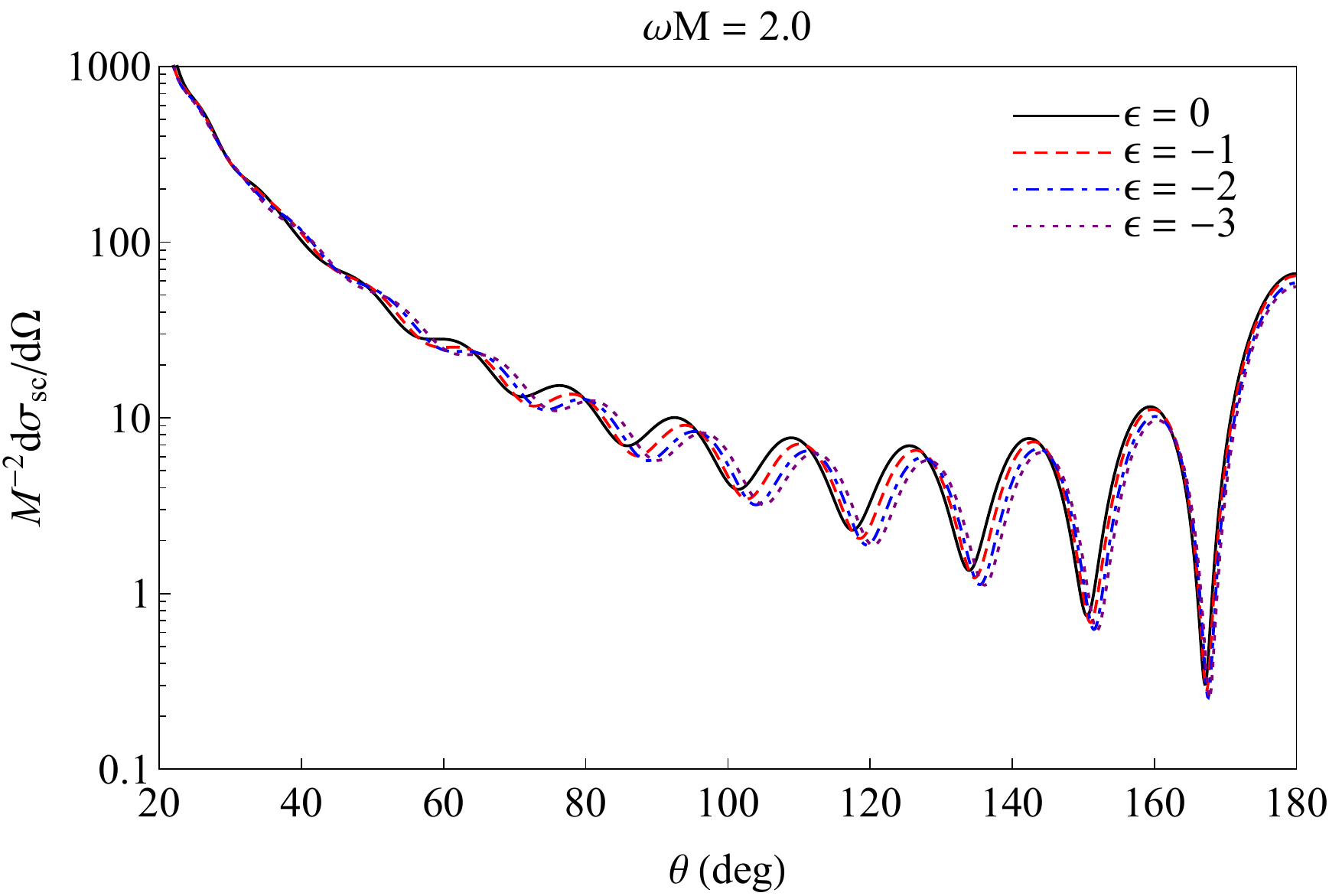}}\\
\subfloat{\label{sfig:a2}\includegraphics[width=\columnwidth]{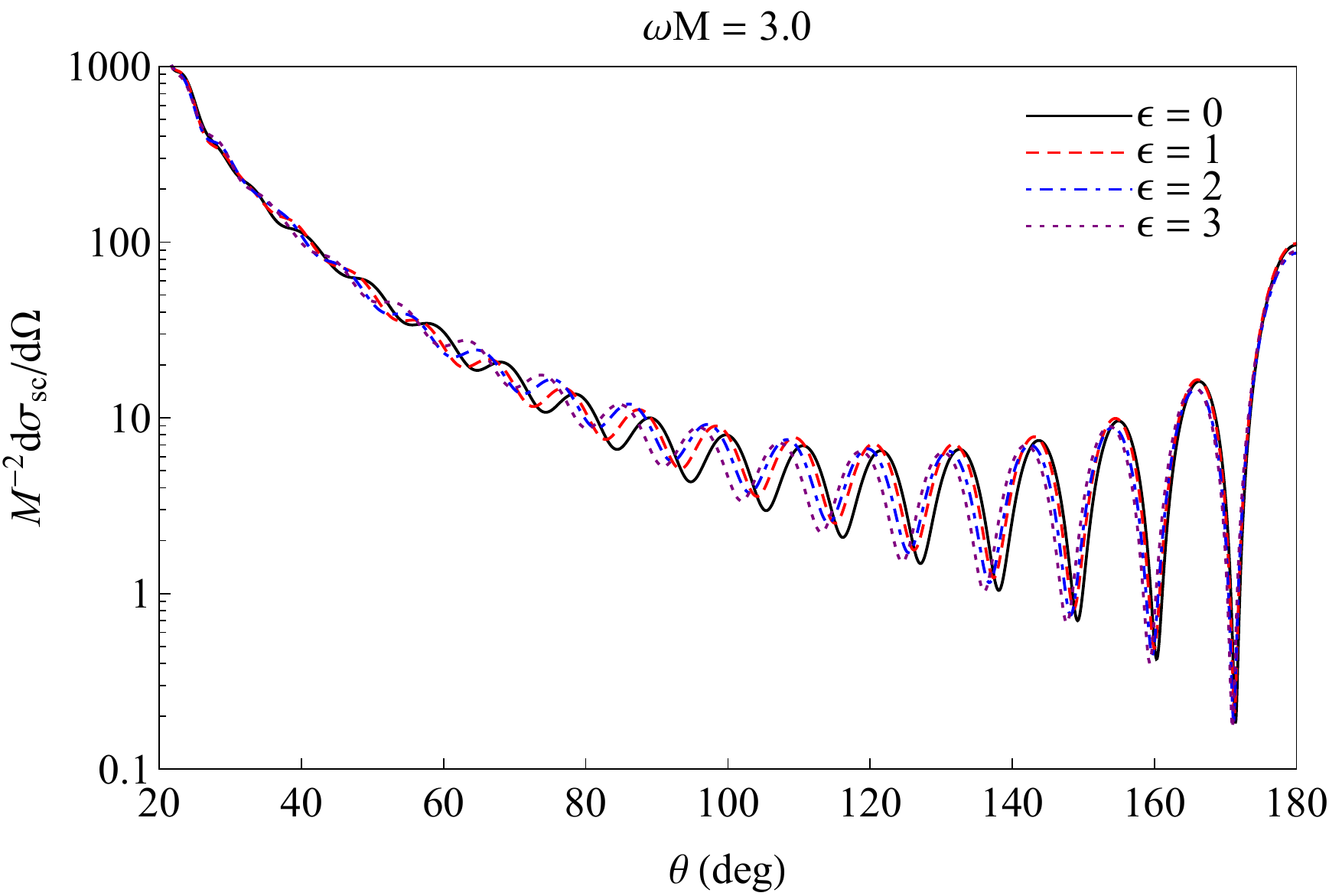}}
\subfloat{\label{sfig:b2}\includegraphics[width=\columnwidth]{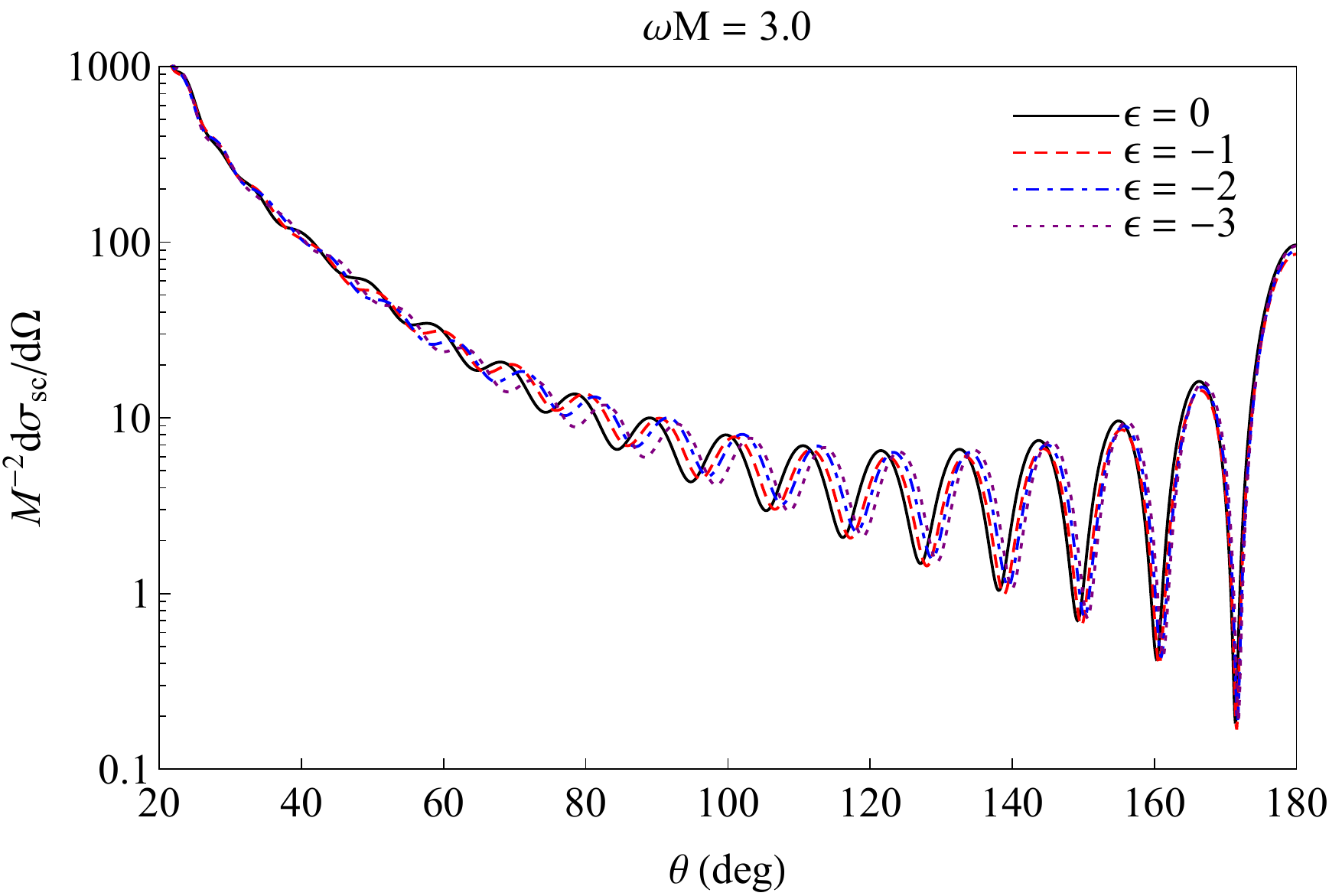}}
\caption{Differential scattering cross section for JPBHs with non-negative (left column) and nonpositive (right column) values of the parameter $\epsilon$, for three choices of frequency, namely $\omega M=1$ (top), $2$ (middle) and $3$ (bottom). The fringe width diminishes as the parameter $\epsilon$ decreases.}
\label{fig:scattering_diff_cross}
\end{figure*}

In Fig.~\ref{fig:scattering_diff_cross}, we plot the differential scattering cross section of JPBHs for the massless scalar field with three choices of frequency, namely $M\omega=1.0$, 2.0 and 3.0, for non-negative (left column) and nonpositive (right column) values of $\epsilon$. We notice that, for a given value of the frequency, the fringe widths (the distance between two consecutive local maxima or minima) get narrower as the value of the parameter $\epsilon$ decreases. Moreover, for positive (negative) values of $\epsilon$, the fringes are shifted to the left (right), as one increases (diminishes) the value of the deformation parameter. 
This feature can be understood considering the monotonic behavior of $b_g$ as a function of the deformation parameter~(see Fig.~\ref{fig:gloryparameters}), since the width of the fringes is proportional to $1/b_g$, in the semiclassical picture. In the semiclassical analysis, the oscillations presented by the differential scattering cross sections are understood as being due to the interference between rays orbiting the BH in opposite senses. 

\begin{figure*}
\centering
\subfloat{\label{sfig:a3}\includegraphics[width=\columnwidth]{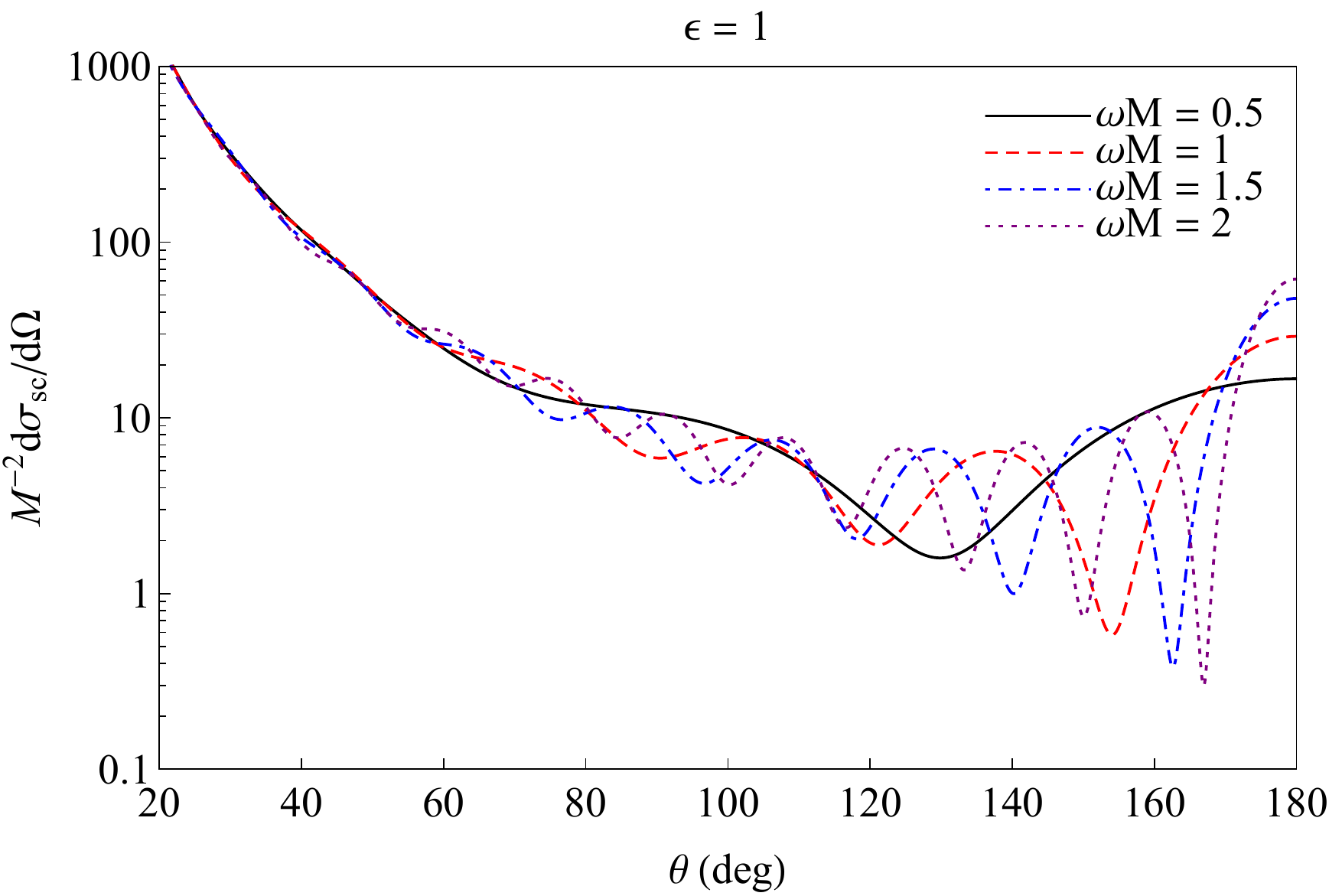}}
\subfloat{\label{sfig:b3}\includegraphics[width=\columnwidth]{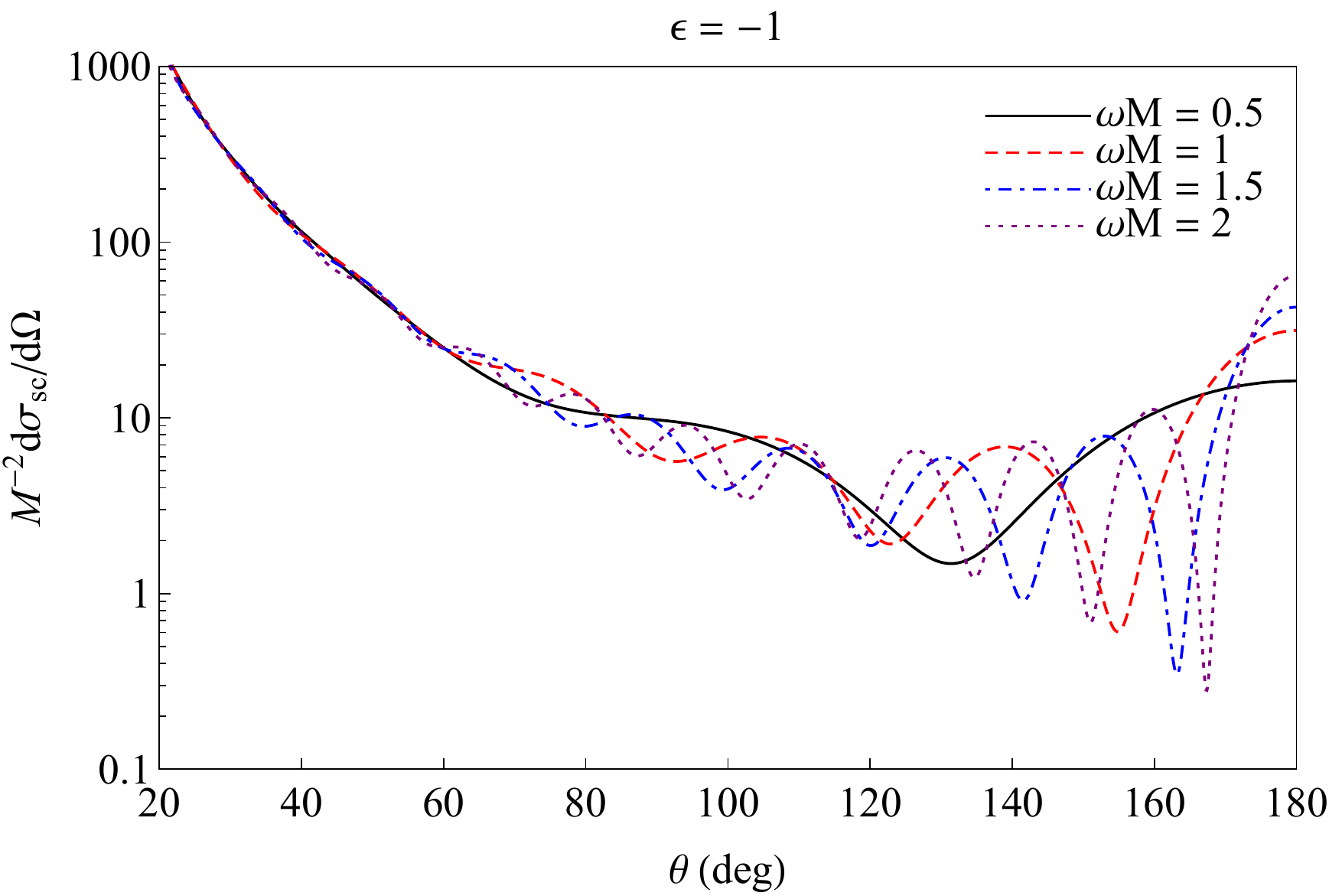}}\\
\subfloat{\label{sfig:a4}\includegraphics[width=\columnwidth]{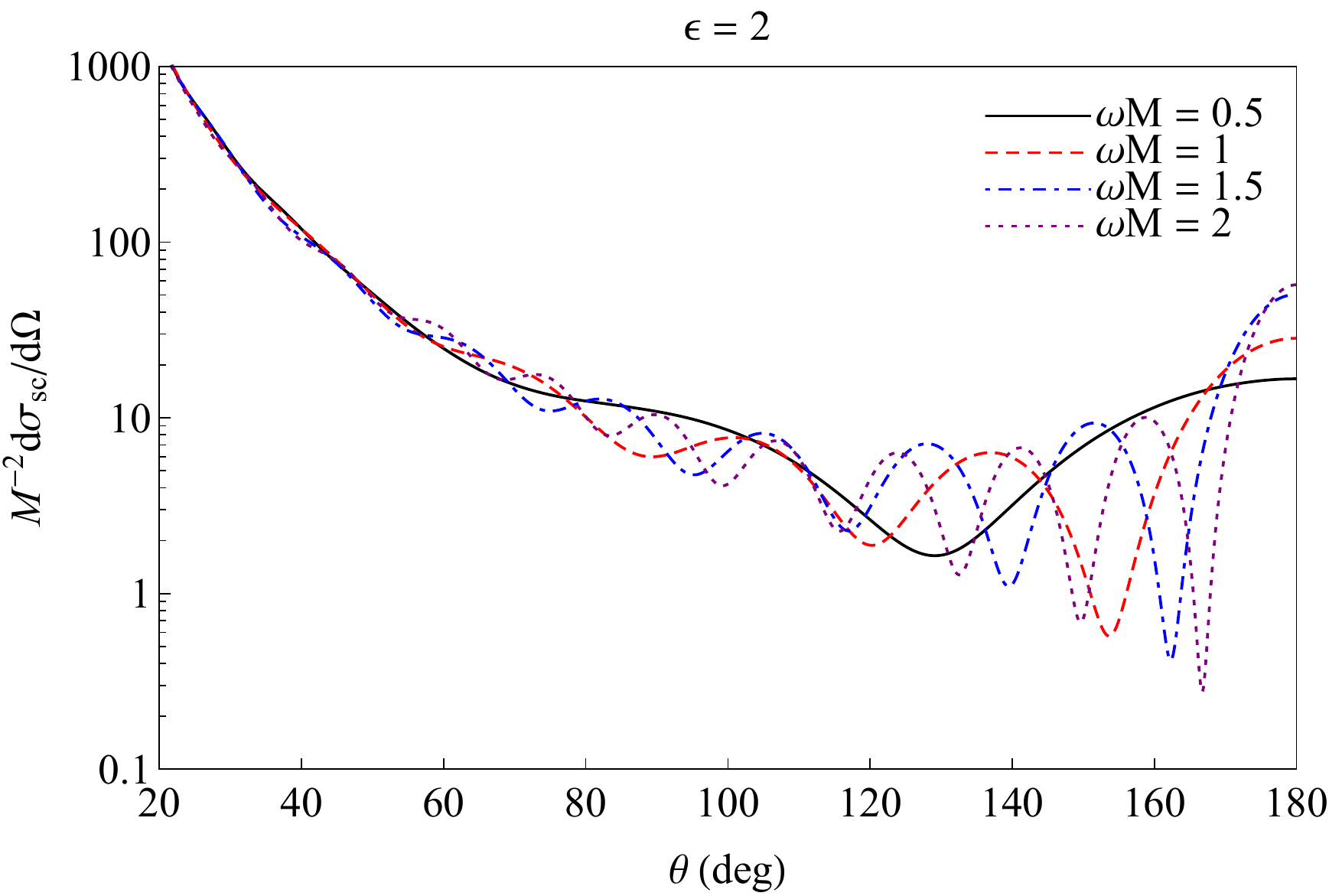}}
\subfloat{\label{sfig:b4}\includegraphics[width=\columnwidth]{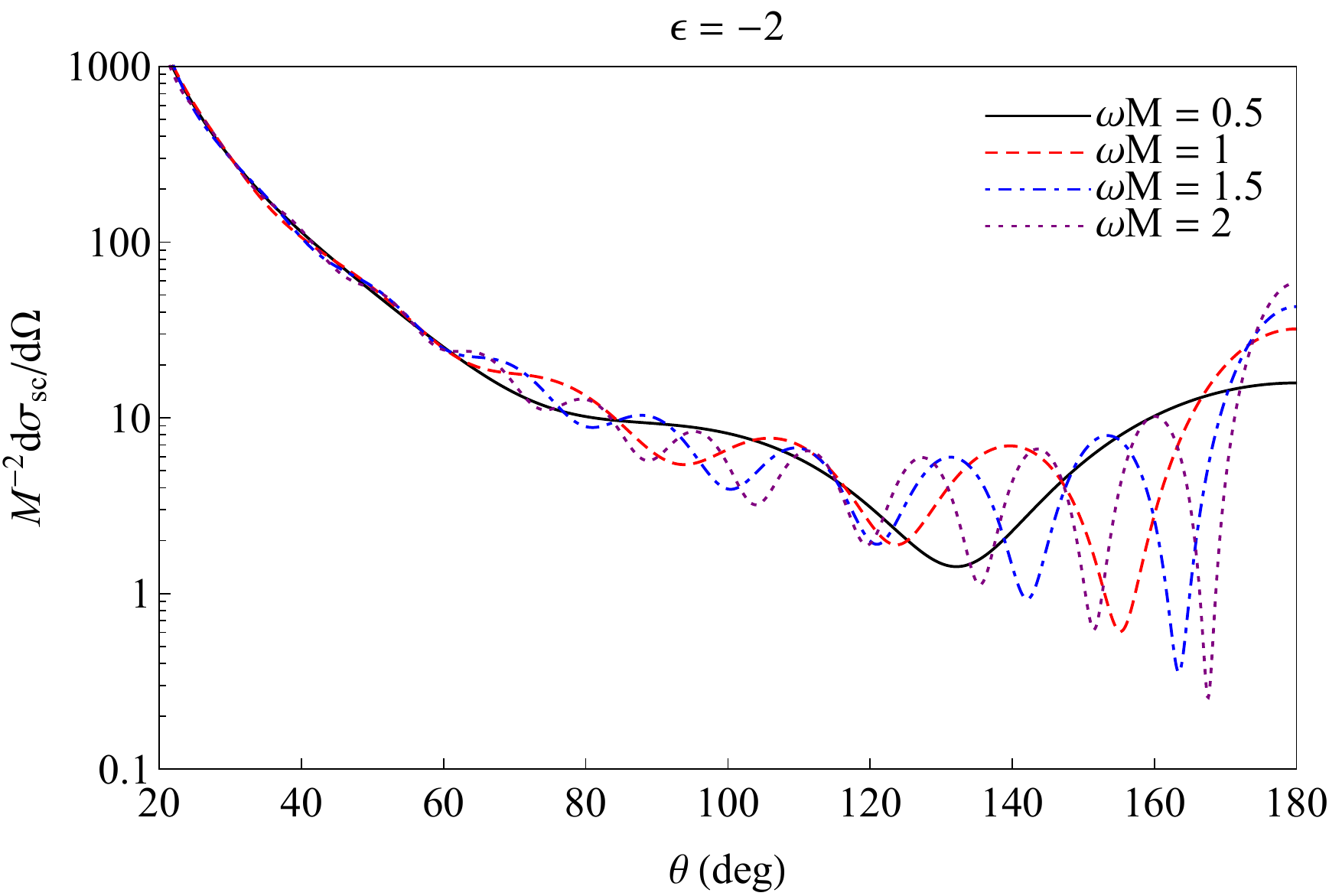}}
\caption{Differential scattering cross section of JPBHs for fixed values of the parameter $\epsilon$, considering four different values of the frequency of the incident scalar wave.}
\label{fig:scattering_diff_cross_omega}
\end{figure*}

In Fig.~\ref{fig:scattering_diff_cross_omega}, we plot the differential scattering cross section of JPBHs for the massless scalar field, considering four choices of the parameter $\epsilon$, namely $\epsilon=-2,-1,1,2$; and different values of the frequency~$\omega$. We notice that, as we increase the value of the frequency, the number of fringes increases and they get narrower. This can be understood considering the (semiclassical) glory approximation~\eqref{eq:glory}, where we see that the argument of the Bessel function is proportional to $\omega$ so that the fringe width is proportional to $1/\omega$.

\begin{figure}[!h]
\includegraphics[width=\columnwidth]{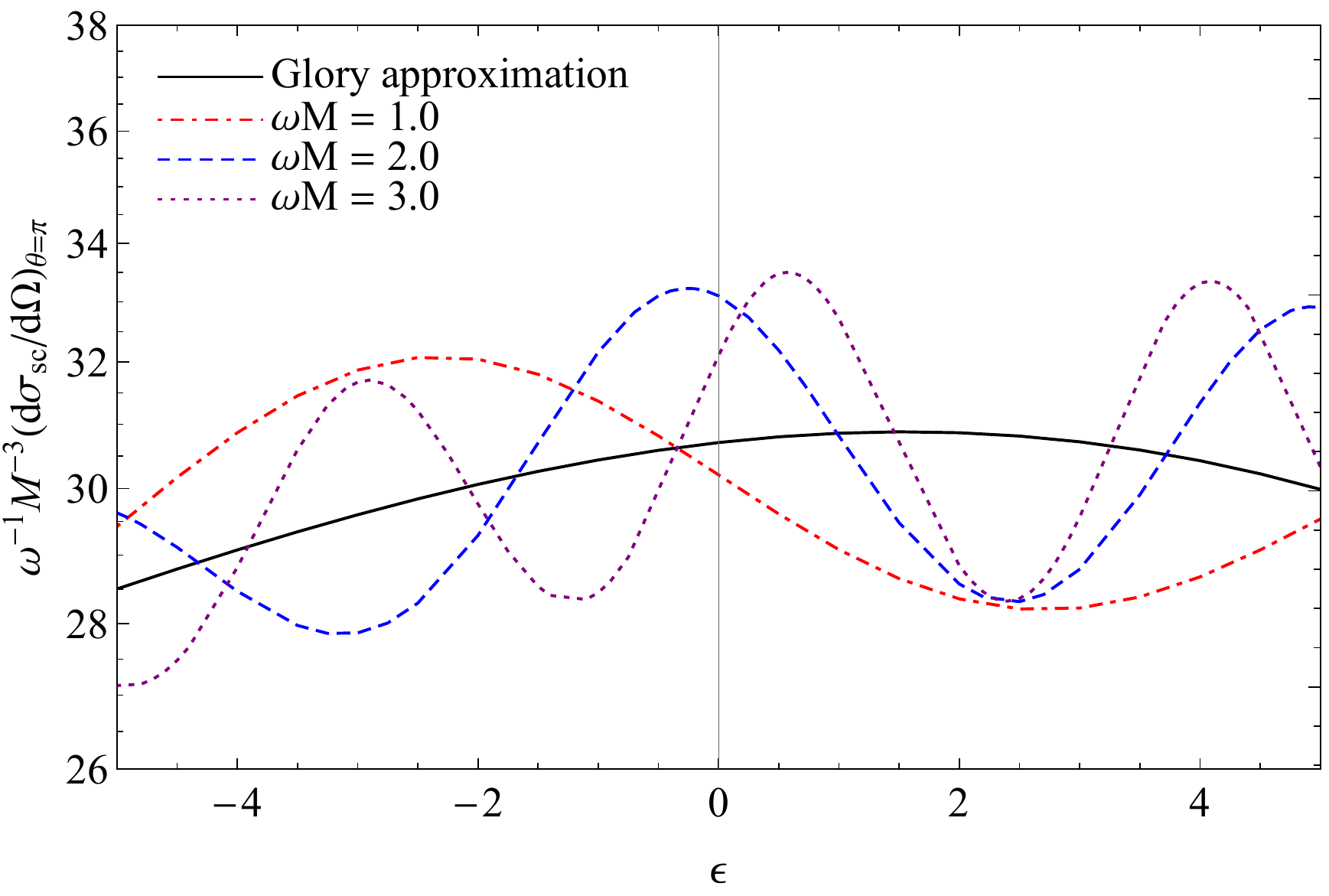}
\caption{Glory peak ($\theta = \pi$) as a function of the deformation parameter $\epsilon$, obtained by the partial waves approach, for $\omega M = 1.0,$ 2.0 and 3.0. We notice that the numerical results oscillate around the glory approximation (solid line).}%
		\label{fig:gloryamp}%
\end{figure}

It is worth pointing out that, although at large scattering angles ($\theta\approx\pi$), a very nice agreement between the glory approximation and the partial waves approach is observed, there is a difference between the magnitude of the peak located at $\theta=\pi$, obtained by these two different approaches. In Fig.~\ref{fig:gloryamp}, this difference is illustrated. From Fig.~\ref{fig:gloryamp}, we notice that the glory peak at $\theta = \pi$, obtained by the partial waves approach, oscillates around the glory amplitude obtained by the semiclassical approximation. This means that the semiclassical approach can help us to interpret the scattering, mainly near the backward direction, but it is not sufficiently accurate to give the magnitude of the differential scattering cross section for arbitrary values of $\epsilon$ and $\omega M$.
\section{Final remarks}\label{sec:remarks}
BHs are natural candidates to test GR, mainly in the strong field regime. Since alternative theories of gravity may arise to amend GR, the analysis of BH solutions in those theories is justified. 
One may consider BH solutions with parametric deviations from the standard BH solutions within GR, constructed without assuming a specific gravity theory.

We have presented the formalism needed to compute the absorption and scattering of massless scalar waves by Schwarzschild-like BHs with parametric deformations and focused in the parametrization proposed by Johannsen and Psaltis~\cite{JP2011}.
In particular, we studied the scattering problem using the partial waves method to obtain our numerical results for arbitrary values of the frequency of the incident wave and of the scattering angle. 
We also considered a classical analysis, based in the null geodesic description, as well as a semiclassical analysis, which is suitable for high frequencies and scattering angles close to the backward direction.
We presented numerical results for JPBHs with one nonvanishing deformation parameter, namely $\epsilon$. We compared our results with the ones for the scattering of massless scalar waves by a Schwarzschild BH in order to infer how the deformation parameter influences the scattering process.

Regarding the scalar absorption by JPBHs, as discussed in Ref.~\cite{MLC:2020EPJC}, the role of the deformation parameter $\epsilon$ is clear: The presence of positive (negative) deformation parameter attenuates (strengthens) the absorption compared with the Schwarzschild case. More specifically, we notice that the increase of the value of the deformation parameter decreases the total absorption cross section. In the high-frequency regime, the total absorption cross section oscillates around the geometrical absorption cross section (obtained via the geodesic analysis), which is confirmed by the sinc approximation~\cite{folacci}. On the other hand, at the low-frequency regime, the absorption cross section tends to the JPBH horizon area, which has the same value as the Schwarzschild BH area with the same mass. This is in agreement with the general result of Refs.~\cite{das, Higuchi:2001}.

In the scattering process, the role of the deformation parameter $\epsilon$ is more involved. When considering small scattering angles, the effects of $\epsilon$ are less relevant, and our numerical results for the differential scattering cross section are well approximated by the classical differential scattering cross section, obtained via the geodesic analysis, and near the forward direction the differential scattering cross section diverges. Near to the backward direction, i.e., for scattering angles close to $\pi$, the role played by the deformation parameter in the wave scattering is more relevant. By analyzing the lightlike trajectories, we found that the value of the glory impact parameter, $b_g$, decreases as the deformation parameter $\epsilon$ increases. Taking into account the behavior of $b_g$, by analyzing the glory scattering formula, we conclude that the scattering fringes get thinner for smaller values of $\epsilon$, which is confirmed by our numerical data. We also found that, as one enhances the values of the deformation parameter, the glory peak obtained via the numerical method oscillates around the glory amplitude obtained via the semiclassical formula. 
Our analysis can be generalized to include additional deformation parameters of the JPBHs.

Scattering processes play an important role in BH physics; since by analyzing the scattered flux of fundamental fields one can, in principle, determine the BH parameters. In particular, through the model independent framework of parametrized black holes, one can use numerical tools to study the scattering phenomena in modified gravity theories. In this context, the scattering by parametrized BHs is a good approach to better understand how nontrivial additional parameters influence BH environments. We hope that future experiments together with scattering investigations can be used for further tests of the no-hair paradigm.

\begin{acknowledgments}
The authors would like to acknowledge 
Funda\c{c}\~ao Amaz\^onia de Amparo a Estudos e Pesquisas (FAPESPA), 
Conselho Nacional de Desenvolvimento Cient\'ifico e Tecnol\'ogico (CNPq)
 and Coordena\c{c}\~ao de Aperfei\c{c}oamento de Pessoal de N\'ivel Superior (CAPES) -- Finance Code 001, from Brazil, for partial financial support. This research has also received funding from the European Union's Horizon 2020 research and innovation programme under the H2020-MSCA-RISE-2017 Grant No. FunFiCO-777740. L. C. S. L. would like to acknowledge IFPA -- Campus Altamira for the support.
\end{acknowledgments}
{}
\end{document}